\documentclass[preprint,aps,11pt,showpacs,nofootinbib,tightenlines]{revtex4}

\usepackage[a4paper,hmarginratio=1:1,vmarginratio=2:3,
totalwidth=16cm,totalheight=24cm]{geometry}

\usepackage{dcolumn}
\usepackage{amsmath}
\usepackage{amssymb}
\usepackage{amsfonts}
\usepackage{epsfig}
\usepackage{graphics}
\usepackage{dcolumn}
\usepackage{longtable}

\newcounter{oldcounter}
\addtocounter{equation}{1}
\begin{document}

\thispagestyle{empty} \vspace{3cm}

\begin{center}
{\Large \textbf{Charmless $B_s\to PP, PV, VV$ Decays Based on the Six-quark Effective Hamiltonian with Strong Phase Effects II}}
\bigskip

\vspace{0.5cm}
Fang Su$^{\ast\dagger}$, Yue-Liang Wu$^{\ast}$,  Yi-Bo Yang$^{\ast\ddagger}$, Ci Zhuang$^{\ast}$\\
\vspace{0.5cm}
$^\ast$ State Key Laboratory of Theoretical Physics \\
Kavli Institute for Theoretical Physics China \\
Institute of Theoretical Physics, Chinese Academy of Sciences, Beijing 100190, China \\
$^{\dagger}$ Institute of Particle Physics, Huazhong Normal University, Wuhan, Hubei, 430079, China\\
$^\ddagger$ Institute of High Energy Physics, Chinese Academy of Sciences, Beijing 100048, China

\end{center}

\medskip
\begin{abstract}
\noindent We provide a systematic study of charmless $B_s \to PP, PV, VV$ decays~($P$ and $V$ denote pseudoscalar and vector mesons, respectively) based on an approximate six-quark operator effective Hamiltonian from QCD. The calculation of the relevant hard-scattering kernels is carried out, the resulting transition form factors are consistent with the results of QCD sum rule calculations. By taking into account important classes of power corrections involving ``chirally-enhanced'' terms and the vertex corrections as well as weak annihilation contributions with non-trivial strong phase, we present predictions for the branching ratios and CP asymmetries of $B_s$ decays into PP, PV and VV final states, and also for the corresponding polarization observables in VV final states. It is found that the weak annihilation contributions with non-trivial strong phase have remarkable effects on the observables in the color-suppressed and penguin-dominated decay modes. In addition, we discuss the SU(3) flavor symmetry and show that the symmetry relations are generally respected.
\end{abstract}


\maketitle

\newpage
\setcounter{page}{1}
\def\thefootnote{\arabic{footnote}}
\setcounter{footnote}{0}

\section{Introduction}

The study of hadronic charmless bottom-meson decays can provide not only an interesting avenue to understand the CP violation and the flavor mixing of quark sector in the Standard Model (SM), but also a powerful means to probe different new physics scenarios beyond the SM~\cite{Antonelli:2009ws,Buchalla:2008jp}. In the past decade, nearly 100 charmless decays of $B_{u,d}$ mesons have been observed at the two B factories, BaBar and Belle. The experimental program to study hadronic $B_s$ decays has also started, with first measurements for the branching ratios of $\bar B_s \to \phi\phi$, $\bar B_s \to K^+\pi^-$ and $\bar B_s \to K^+K^-$ made available by the CDF Collaboration~\cite{Acosta:2005eu,Abulencia:2006psa,Morello:2006pv}. Remarkably, the first evidence for direct CP violation involving $\bar B^0_s \to K^+\pi^-$ and its CP conjugate mode has also been reported by the CDF Collaboration, with $A_{CP}(\bar B_s \to K^+\pi^-)=(39\pm 15\pm 8)\%$~\cite{Morello:2006pv}. Because of the possibilities of new discoveries, the $B_s$ system will be the main focus of the forthcoming experiments at Fermilab, LHCb and Super-B factories.

Theoretically, analogous to the $B_{u,d}$-meson decays, the charmless two-body $B_s$-meson decays have also been studied in great detail in the literature. For example, detailed estimates have been undertaken in the framework of generalized factorization~\cite{Tseng:1998wm,Chen:1998dta}, QCD factorization~(QCDF)~\cite{Li:2003hea,Beneke:2003zv,Beneke:2006hg}, perturbative QCD~(pQCD)~\cite{Chen:2001sx, Ali:2007ff} and soft-collinear effective theory~(SCET)~\cite{Williamson:2006hb,Wang:2008rk}. Furthermore, various New Physics~(NP) effects in $B_s$ decays have also been considered in \cite{Bao:2008hd,Baek:2006pb,Xu:2009we}. With the experimental developments at Fermilab, LHC-b and Super-B factories, more and more theoretical studies on $B_s$-meson decays are expected in the forthcoming years.

In this work, we shall reexamine hadronic $B_s$-meson decays based on an approximate six-quark operator effective Hamiltonian, which has been applied to $B_{u,d} \to PP, PV, VV$~(where $P$ and $V$ denote pseudoscalar and vector mesons, respectively) decays~\cite{Su:2008mc,Su:2010vt}. During the evaluation of the hadronic matrix elements of effective six-quark operators, the encountered infrared singularity caused by the gluon exchanging interaction is simply cured by the introduction of an energy scale $\mu_g$, which plays the role of infrared cut-off without violating the gauge invariance, as motivated by the gauge invariant loop regularization method~\cite{LRC1,LRC2,Cui:2008uv}. In general, the infrared cut-off energy scale runs with the final-state mass energy, we adopt $\mu_g\sim500$~MeV in VV final states and $\mu_g\sim350$~MeV in $PP, PV$ final states, due to the fact that the former concerns higher mass energy than the latter.

The calculation of strong phase from non-perturbative effects is a hard task, there exists no efficient approach to evaluate reliably the strong phases. From the study of hadronic $B_{u,d}$ decays in our previous papers~\cite{Su:2008mc,Su:2010vt}, it has been shown that the effective Wilson coefficients and the annihilation contribution with a strong phase lead to a better prediction in our framework to explain the observed branching ratios and CP asymmetries. As for the effective Wilson coefficients associated with the operators, rather than considering only the power corrections to the color-suppressed tree topology~\cite{Cheng:2009mu}, we also attribute possible non-perturbative corrections parameterized by $\widetilde{V_1}$ and $\widetilde{V_2}$ to the corresponding two operator structures $(V-A)\times(V-A)$ and $(V-A)\times(V+A)$.

It is interesting to note that, with the above simplified prescriptions, the method developed based on the six-quark effective Hamiltonian allows us to calculate the relevant B to light meson transition form factors, and the resulting predictions are consistent with the results of light-cone QCD sum rules. To further test the feasibility of our framework, we shall extend our method to charmless $B_s \to PP, PV, VV$ decays. It is seen that the predictions for the branching ratios in the tree-dominated $B_s$ decays are generally in good agreement among different theoretical approaches, while there might be big discrepancies in color-suppressed, penguin-dominated and annihilation-dominated $B_s$ decays: in QCDF approach~\cite{Cheng:2009mu}, it favors big color-suppressed and penguin-dominated contributions, while in pQCD approach~\cite{BsPV}, it prefers a big annihilation contribution. In our approach, the results stand between the QCDF and pQCD. With this situation, it is expected that the future precise experimental datas will give us an unambiguous answer. In addition, we discuss the SU(3) flavour symmetry in the decays which have been experimentally observed, and show that in some other interesting decay channels the symmetry relations are generally respected.

Our paper is organized as follows. In Section~\ref{sec:sqeh}, firstly, we briefly review the primary six-quark diagrams with the exchanges of a single W-boson and a single gluon, as well as the corresponding initial six-quark operators. Then we present the treatments of the singularities caused by the gluon exchanging interactions and the on mass-shell fermion propagator, as well as the vertex corrections and annihilation contributions. Section~\ref{Input} contains all the input parameters used in our calculations. In Section~\ref{sec:fitting}, we give our numerical predictions and discussions for $B_s\to PP, PV, VV$ decays. Our conclusions are presented in the last section. Some details on the decay amplitudes are given in the Appendix.

\section{Theoretical Framework}\label{sec:sqeh}

\subsection{Four-Quark Operator Effective Hamiltonian}

Let us start from the four-quark effective operators in the effective weak Hamiltonian. The initial four-quark operator due to weak interaction via W-boson exchange is given as follows for the B-meson decays:
\begin{equation}
O_{1}=(\bar{q}^u_{i}b_{i})_{V-A}(\bar{q}^d_{j}u_{j})_{V-A}, \qquad
q^u=u,\ c, \quad  q^d = d,\ s.
\end{equation}
The complete set of four-quark operators are obtained from QCD and QED corrections which contain the gluon-exchange diagrams, strong penguin diagrams and electroweak penguin diagrams. The resulting effective Hamiltonian~(for $b\to s$ transition) with four-quark operators is given as follows~\cite{4qham}:
\begin{eqnarray}
H_{\rm eff}\, =\, {\frac{G_F}{\sqrt{2}}} \sum_{q=u,c}
%\lambda_q^{s(d)}
\lambda_q^{s} \left[C_1(\mu)O_1^{(q)}(\mu) +C_2(\mu)O_2^{(q)}(\mu)+
\sum_{i=3}^{10}C_i(\mu)O_i(\mu)\right]+{\rm h.c.}\;,\label{eq:hpk}
\end{eqnarray}
where $\lambda_q^{s} = V_{qb}V^*_{qs}$ are products of the CKM matrix elements, $C_i(\mu)$ the Wilson coefficient functions~\cite{4qham}, and $O_i(\mu)$ the four-quark operators:
\begin{eqnarray}
\begin{array}{ll}
\displaystyle O_1^{(q)}\, =\,
(\bar{q}_ib_i)_{V-A}(\bar{s}_jq_j)_{V-A}\;, & \displaystyle
O_2^{(q)}\, =\,(\bar{s}_ib_i)_{V-A}(\bar{q}_jq_j)_{V-A}\;,
\\
\displaystyle O_3\,
=\,(\bar{s}_ib_i)_{V-A}\sum_{q'}(\bar{q}'_jq'_j)_{V-A}\;,
&\displaystyle O_4\,
=\,\sum_{q'}(\bar{q}'_ib_i)_{V-A}(\bar{s}_jq'_j)_{V-A}\;,
\\
\displaystyle O_5\,
=\,(\bar{s}_ib_i)_{V-A}\sum_{q'}(\bar{q}'_jq'_j)_{V+A}\;,
&\displaystyle O_6\,
=\,(\bar{s}_i b_j)_{V-A}\sum_{q}(\bar{q}_j q_i)_{V+A}\;,
\\
\displaystyle O_7\,
=\,\frac{3}{2}(\bar{s}_ib_i)_{V-A}\sum_{q'}e_{q'}(\bar{q}'_jq'_j)_{V+A}\;,&
\displaystyle O_8\, =\,
\frac{3}{2}(\bar{s}_i b_j)_{V-A}\sum_{q}e_{q}
(\bar{q}_j q_i)_{V+A}\;,
\\
\displaystyle O_9\,
=\,\frac{3}{2}(\bar{s}_ib_i)_{V-A}\sum_{q'}e_{q'}(\bar{q}'_jq'_j)_{V-A}\;,&
\displaystyle O_{10}\, =\,
\frac{3}{2}\sum_{q'}e_{q'}(\bar{q}'_ib_i)_{V-A}(\bar{s}_jq'_j)_{V-A}\;.\\
\end{array}
\label{eq:o}
\end{eqnarray}
Here the Fermi constant $G_F=1.16639\times 10^{-5}\;{\rm GeV}^{-2}$, $(\bar{q}'q')_{V\pm A} = \bar q' \gamma_\mu (1\pm \gamma_5)q'$, and $i, j$ are the color indices. The index $q'$ in the summation of the above operators runs through $u,\;d,\;s$, $c$, and $b$. The effective Hamiltonian for the $b\to d$ transition can be obtained by changing $s$ into $d$ in Eqs.~(\ref{eq:hpk}) and (\ref{eq:o}).

\subsection{Six-quark Diagrams and Effective Operators}\label{sec:sqd}

\begin{figure}[htbp]
\begin{center}
\includegraphics[scale=0.50]{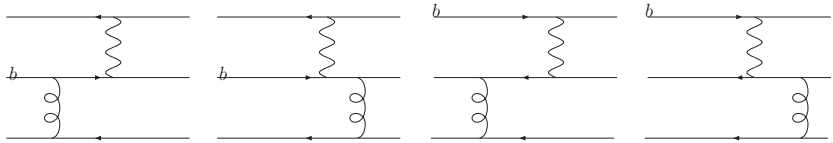}\\
  \caption{\small{Four different six-quark diagrams with a single W-boson exchange and a single gluon exchange.}}\label{pic:4insert}
  \end{center}
\end{figure}

As mesons are regarded as quark and anti-quark bound states, the two-body hadronic B-meson decays actually involve three quark-antiquark pairs. It is then natural to consider the six-quark Feynman diagrams which lead to three effective quark-antiquark currents. The initial six-quark diagrams of weak decays contain one W-boson exchange and one gluon exchange, thus there are four different diagrams as the gluon exchange interaction can occur for each of four quarks in the W-boson exchange diagram, see Fig.~\ref{pic:4insert}.

%%%%%%%%%%%%%%%%%
%%%%%%%%%%%%%%%%%%%%%

The resulting initial effective operators contain four terms corresponding to the four diagrams, respectively. In a good approximation, the four quarks via W-boson exchange can be regarded as a local four-quark interaction at the energy scale much below the W-boson mass, while the two QCD vertices due to gluon exchange are at two independent space-time points, the resulting effective six-quark operators are in general non-local. The six-quark operators corresponding to the four diagrams in Fig.~\ref{pic:4insert} are found to be
\begin{eqnarray}
  O^{(6)}_{q_1}\, &=&\, 4\pi\alpha_s \int\!\!\int  \frac{\emph{d}^4k}{(2\pi)^4}\, \frac{\emph{d}^4p}{(2\pi)^4}\,e^{-i((x_1-x_2)p+(x_2-x_3)k)}
  (\bar{q'}(x_3)\gamma_{\nu}T^{a} q'(x_3))\frac{1}{k^2+i\epsilon}\nonumber\\
  &&\times(\bar{q}_{2}(x_1)\Gamma_{1}\frac{p\!\!\!/+m_b}{p^2-m_b^2+i\epsilon}\gamma^{\nu}T^{a} q_{1}(x_2))*
  (\bar{q}_{4}(x_1) \Gamma_{2} q_{3}(x_1)),\nonumber\\[0.2cm]
  O^{(6)}_{q_2}\, &=&\, 4\pi\alpha_s \int\!\!\int  \frac{\emph{d}^4k}{(2\pi)^4}\, \frac{\emph{d}^4p}{(2\pi)^4}\,e^{-i((x_1-x_2)p+(x_2-x_3)k)}
  (\bar{q'}(x_3)\gamma_{\nu}T^{a} q'(x_3))\frac{1}{k^2+i\epsilon}\nonumber\\
  &&\times(\bar{q}_{2}(x_2)\frac{p\!\!\!/+m_{q_2}}{p^2-m_{q_2}^2+i\epsilon}\gamma^{\nu}T^{a}\Gamma_{1} q_{1}(x_1))*
  (\bar{q}_{4}(x_1) \Gamma_{2} q_{3}(x_1)),\nonumber\\[0.2cm]
  O^{(6)}_{q_3}\, &=&\, 4\pi\alpha_s \int\!\!\int  \frac{\emph{d}^4k}{(2\pi)^4}\, \frac{\emph{d}^4p}{(2\pi)^4}\,e^{-i((x_1-x_2)p+(x_2-x_3)k)}
  (\bar{q'}(x_3)\gamma_{\nu}T^{a} q'(x_3))\frac{1}{k^2+i\epsilon}\nonumber\\
  &&\times(\bar{q}_{2}(x_1)\Gamma_{1} q_{1}(x_1))*
  (\bar{q}_{4}(x_1) \Gamma_{2} \frac{p\!\!\!/+m_{q_3}}{p^2-m_{q_3}^2+i\epsilon}\gamma^{\nu}T^{a} q_{3}(x_2)),\nonumber\\[0.2cm]
  O^{(6)}_{q_4}\, &=&\, 4\pi\alpha_s \int\!\!\int  \frac{\emph{d}^4k}{(2\pi)^4}\, \frac{\emph{d}^4p}{(2\pi)^4}\,e^{-i((x_1-x_2)p+(x_2-x_3)k)}
  (\bar{q'}(x_3)\gamma_{\nu}T^{a} q'(x_3))\frac{1}{k^2+i\epsilon}\nonumber\\
  &&\times(\bar{q}_{2}(x_1)\Gamma_{1} q_{1}(x_1))*
  (\bar{q}_{4}(x_2)\frac{p\!\!\!/+m_{q_4}}{p^2-m_{q_4}^2+i\epsilon}\gamma^{\nu}T^{a} \Gamma_{2} q_{3}(x_1)),
\label{eq:six}
\end{eqnarray}
where $k$ and $p$ correspond to the momenta of gluon and quark in their propagators. $q_1$ is usually set to be the heavy b quark. $x_1$, $x_2$ and $x_3$ are space-time points corresponding to the three vertices. The color index is summed between $q_1, q_2$ and $q_3, q_4$. Note that all the six-quark operators are proportional to the QCD coupling constant $\alpha_s$ due to gluon exchange. Thus the initial six-quark operator is given by summing over the above four operators:
\begin{eqnarray}
  O^{(6)}=\sum_{j=1}^4 O^{(6)}_{q_j}.
\end{eqnarray}
Unlike the classical four-quark effective operator, the six-quark operators used here are non-local with quark and gluon propagators inserted into them.

Before proceeding, we would like to address that the above effective six-quark operators obtained with the gluon exchanging diagrams are similar to the four fermion interactions in the Nambu-Jona-Lasino model. As it is well known that the effective four quark interactions proposed in Nambu-Jona-Lasino model can well lead to the dynamical chiral symmetry breaking via quark condensate, which was motivated from a single gluon exchange operator.The nonperturbative effects were taken into account from the gluon coupling constant $\alpha_s$ by running it down to the low energy scale, and from the QCD confinement scale $\mu \sim \Lambda_{QCD} \simeq 300\sim 400$ MeV which may be regarded as a dynamical gluon mass scale in the infrared region:
\begin{equation}
L = \frac{\alpha_s}{\mu^2} \bar{\psi} \psi \bar{\psi} \psi \nonumber
\end{equation}

Such a picture has successfully described the nonperturbative effects of QCD at low energy dynamics. We all believe that perturbative QCD cannot correctly deal with the hadronic decays, which is actually the main motivation in our paper to develop an alternative approach to treat the hadronic two-body decays within the framework of QCD. It is clear that QCDF approach cannot compute theoretically the hadronic amplitudes from the framework of QCD and it needs to have some inputs for the form-factors of hadronic matrix elements. It is well known that in the calculations of hadronic decay amplitudes based on the effective four-quark Hamiltonian, the perturbative contributions of QCD are characterized by the Wilson coefficient functions running from high energy scales to low energy scale around $\mu \simeq 1.0\sim 1.5$ GeV, and the nonperturbative contributions are carried out by evaluating the hadronic matrix elements of the effective quark operators at low energy,  where some nonperturbative effects are considered in the wave functions of hadrons. In our approach with non-local effective six-quark Hamiltonian, the treatment is similar. The perturbative contributions are included in the Wilson coefficient functions with an additional QCD coupling constant $\alpha_s$ due to the gluon exchanging diagram for obtaining effective six-quark operators,  and the gluon-gluon interactions are partially taken into account in the running of $\alpha_s$ from high energy scale to low energy scale around $\mu \simeq 1.5$ GeV. The nonperturbative contributions are considered in the evaluation of hadronic matrix elements of nonlocal effective six-quark operators, where the additional nonperturbative effects are taken into account in the non-local effective quark operators and the QCD confinement scale $\mu \sim \Lambda_{QCD} \sim 300-400$ MeV due to gluon exchanging diagram. The strong gluon-gluon interactions at low energy scale are effectively characterized by a dynamical gluon mass or infrared cut-off scale due to QCD confinement, which is similar to the Nambu-Jona-Lasino model.

With the above considerations, the QCD factorization approach with six-quark operator effective Hamiltonian enables us to evaluate all the hadronic matrix elements of two-body hadronic B-meson decays. For the hadronic matrix elements relevant to $B_s\to PP, PV, VV$ decays, we shall list them in the Appendix.

\subsection{Treatment of Singularities}\label{sec:TOD}

In the evaluation of hadronic matrix elements, there are two kinds of singularities. One singularity stems from the infrared divergence of gluon exchanging interaction, and the other one from the on mass-shell divergence of internal quark propagator.

In general, a Feynman diagram will yield an imaginary part for the decay amplitudes when the virtual particles in the diagram become on mass-shell, and the resulting diagram can be considered as a genuine physical process. It is well known that, when applying the Cutkosky rule~\cite{cutkosky} to deal with a physical-region singularity of all propagators, the following formula holds:
\begin{eqnarray}
\frac{1}{p^2-m_q^2+i\epsilon}=P\biggl[\frac{1}{p^2-m_q^2}
\biggl]-i\pi\delta[p^2-m_q^2],\label{quarkd}
\end{eqnarray}
which is known as the principal integration method, and the integration with the notation of capital letter $P$ is the so-called principal integration.

However, the Cutkosky rule may not directly be used to treat the singularities from the infrared divergence of massless gluon propagator and also the light-quark propagators due to the confinement and dynamical chiral symmetry breaking of strong interactions. In fact, integration with those propagators is sensitive to the infrared cut-off for gluon and light-quark propagators, and diverge to infinity when the cut-off becomes zero. A modified integration with different parameters for different channels is used in QCDF framework~\cite{Beneke:2003zv,Cheng:2009mu}, while the transverse momentum $k_T$ dominating in the zero momentum fraction is added to the propagator in pQCD framework~\cite{lihn}. In our approach, we prefer to introduce the cut-off energy scales for both gluon and light quark motivated from the symmetry-preserving loop regularization\cite{LRC1,LRC2,Cui:2008uv} and in order to investigate the infrared cut-off dependence for the theoretical predictions:
\begin{eqnarray}
   \frac{1}{k^2}\frac{p\!\!/\,+m_q}{(p^2-m_q^2)} &\to&
   \frac{1}{(k^2-\mu_g^2+i\epsilon)}\frac{p\!\!/\,+\mu_q}{(p^2-\mu_q^2+i\epsilon)}~~~(\text{q is a light quark}).
\end{eqnarray}

\begin{figure}[htbp]
\begin{center}
  \includegraphics[scale=0.6]{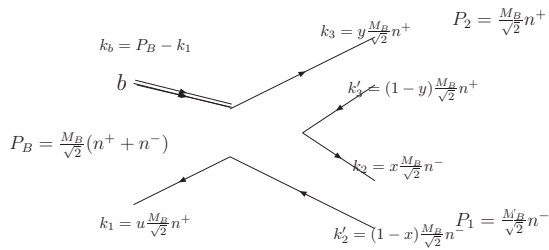}
  \caption{Definition of momenta in $B\rightarrow M_1M_2$ decay. The light-cone coordinate is adopted with $(n^+, n^-, \vec{k}_{\bot})$.} \label{pic:definition}
  \end{center}
\end{figure}

It is noted that, as the gauge dependent term $k_{\mu}k_{\nu}$ can always be rewritten as linear combinations of the momenta $p_{\alpha}$ on the external lines of the spectator quark, which are all on mass-shell in our case~(as defined in Fig.~\ref{pic:definition}), their contributions are equal to zero once the equation of motion is used. Our results are therefore gauge independent.

\subsection{Vertex Corrections and Annihilation Contributions}\label{vetex corrections}

As shown in Ref.~\cite{0508041}, the CP-violating observables may be improved by adding vertex corrections. Furthermore, the vertex corrections were proposed to improve the scale dependence of Wilson coefficients of factorizable emission amplitudes in QCDF~\cite{NPB.606.245}. Those coefficients are always combined as $C_{2n-1}+\frac{C_{2n}}{N_C}$ and $C_{2n}+\frac{C_{2n-1}}{N_C}$, which, after taking into account the vertex corrections, are modified to
\begin{eqnarray}
C_{2n-1}(\mu)+\frac{C_{2n}}{N_c}(\mu) \to
C_{2n-1}(\mu)+\frac{C_{2n}}{N_c}(\mu)
+\frac{\alpha_s(\mu)}{4\pi}C_F\frac{C_{2n}(\mu)}{N_c} V_{2n-1}(M_2)
\;,&&
\nonumber\\[0.2cm]
C_{2n}(\mu)+\frac{C_{2n-1}}{N_c}(\mu) \to
C_{2n}(\mu)+\frac{C_{2n-1}}{N_c}(\mu)
+\frac{\alpha_s(\mu)}{4\pi}C_F\frac{C_{2n-1}(\mu)}{N_c} V_{2n}(M_2) \;,&&
\end{eqnarray}
with $n=1,...,5$, and $M_2$ being the meson emitted from the weak vertex. In the naive dimensional regulation~(NDR) scheme, $V_i(M)$ are given by~\cite{Beneke:2003zv,NPB.606.245}
\begin{eqnarray}
V_i(M) &=&\left\{
\begin{array}{ll}
V_1(M) = 12\ln(\frac{m_b}{\mu})-18+\int_0^1 dx\, \phi_a(x)\, g(x)\;, &
\mbox{\rm for }i=1-4,9,10\;,
\\[0.2cm]
V_2(M) = -12\ln(\frac{m_b}{\mu})+6-\int_0^1dx\, \phi_a(1-x)\, g(1-x)\;, &
\mbox{\rm for }i=5,7\;,
\\[0.2cm]
V_3(M) = -6 +\int_0^1 dx\,\phi_b(x)\, h(x) \;, &
 \mbox{\rm for }i=6,8\;,
\end{array}
\right.\label{vim}
\end{eqnarray}
where $\phi_a(x)$ and $\phi_b(x)$ denote the leading-twist and twist-3 distribution amplitudes for a pseudoscalar or a longitudinally polarized vector meson, respectively. While for a transversely polarized vector final state, $\phi_a(x)=\phi_{\pm}(x,\mu)$ and $\phi_b(x)=0$. The functions $g(x)$ and $h(x)$ used in the integration are given respectively as~\cite{Beneke:2003zv}
\begin{eqnarray}
g(x) &=& 3\left( \frac{1-2x}{1-x}\ln{x} -i\,\pi \right)\nonumber\\
& & +\left[ 2\,{\rm Li}_2(x)-\ln^2 x +\frac{2\ln
x}{1-x}-(3+2i\,\pi)\ln x - (x\leftrightarrow 1-x) \right] \;,\nonumber
\\[0.2cm]
h(x) &=& 2\,{\rm Li}_2(x)-\ln^2 x -(1+2i\,\pi)\ln x -
(x\leftrightarrow 1-x) \;.
\end{eqnarray}

\begin{figure}[t]
\begin{center}
  \includegraphics[scale=0.5]{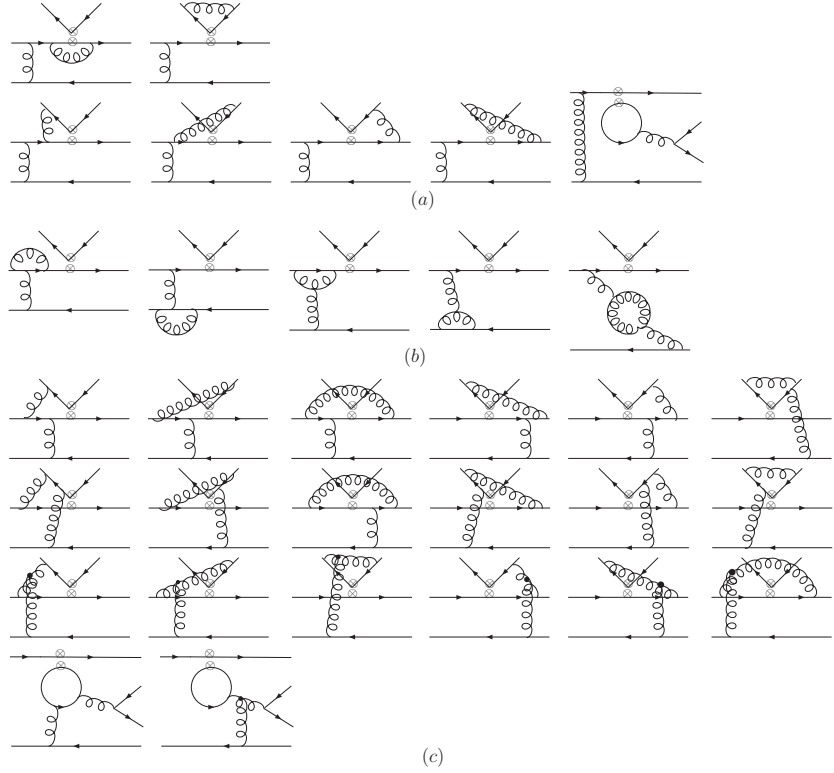}
  \caption{\small The diagrams in (a), (b) and (c) are loop contributions only to the effective weak vertex~(type I), only to the gluon vertex~(type II), and for both weak and strong vertices~(type III), respectively.}\label{pic:loop}
  \end{center}
\end{figure}

To further improve our predictions, we shall examine an interesting case that vertices receive additional large non-perturbative contributions, namely the Wilson coefficients $a_i =C_i+\frac{C_{i\pm1}}{N_C}$ are modified to the following effective ones:
\begin{eqnarray}\label{aeff}
a_i \to a_i^{eff} = C_i(\mu)+\frac{C_{i\pm1}}{N_c}(\mu)
+\frac{\alpha_s(\mu)}{4\pi}C_F\frac{C_{i\pm1}(\mu)}{N_c}\left( V_1(M_2)+\widetilde{V}_1(M_2)\right)
\;, &&(i=1-4,9,10),
\nonumber\\[0.2cm]
a_i \to a_i^{eff} = C_i(\mu)+\frac{C_{i\pm1}}{N_c}(\mu)
+\frac{\alpha_s(\mu)}{4\pi}C_F\frac{C_{i\pm1}(\mu)}{N_c}\left( V_2(M_2)+\widetilde{V}_2(M_2)\right)
\;, &&(i=5,7), \\
a_i \to a_i^{eff} = C_i(\mu)+\frac{C_{i\pm1}}{N_c}(\mu)
+\frac{\alpha_s(\mu)}{4\pi}C_F\frac{C_{i\pm1}(\mu)}{N_c}\left( V_3(M_2)+\widetilde{V}_2(M_2) \right)
\;, &&(i=6,8). \nonumber
\end{eqnarray}
The corrections $\widetilde{V}_1(M_2)$ and $\widetilde{V}_2(M_2)$ depend on whether the meson $M_2$ is a pseudoscalar or a vector meson, and could be caused from the higher order QCD corrections and some non-local effects at low energy scale as shown in Fig.~\ref{pic:loop}. It can be argued from a naive dimensional analysis that a large number of type III diagrams in Fig.3c are in general no longer suppressed at low energy scale and may lead to a significant contribution at low energy scale. While a complete calculation of their effects at low energy scale is not an easy task and beyond the purpose of our present paper, we may first treat them as input parameters and will make a detailed investigation elsewhere. In comparing to the vertex corrections $V_i(M)$ (i=1,2,3), there could also be in general three type of vertex corrections $\widetilde{V}_i(M_2)$. While in our computation, we have only introduced two kinds of vertex corrections $\widetilde{V}_1 (M_2)$ and $\widetilde{V}_2 (M_2)$. This is just for the simplicity of considerations by assuming that the additional two kinds of vertex corrections with strong phases correspond to two kinds of operator structures with current-current interactions $(V-A)\otimes (V-A)$ and $(V-A)\otimes (V+A)$. As it was shown in our previous work\cite{Su:2010vt} that adopting $\widetilde{V}_1(P) = 26 e^{-\frac{\pi}{3} i}$, $\widetilde{V}_2(P)= -26 $, $\widetilde{V}_1(V) = 15 e^{\frac{\pi}{8} i}$, and $\widetilde{V}_2(V)=-15 e^{\frac{\pi}{8} i}$, both the branching ratios and the CP asymmetries of most $B \to PP, PV, VV$ decay modes are improved, we shall take the same input for the $B_s \to PP, PV, VV$ decays.

%%%%%%%%%%%%%%%%%%%%%%%
%%%%%%%%%%%%%%%%%%%%%%%%%%%%

As for the weak annihilation contributions, most of them are from factorizable annihilation diagrams with the $(S-P)\times (S+P)$ effective four-quark vertex:
\begin{eqnarray}\label{annihilation}
A_{SP}^{P_1 P_2}(M) \sim \int dxdy \frac{(\mu_{P_1}+\mu_{P_2})y(1-y)} {(x(1-y)m_B^2-\mu_g^2+i\epsilon)((1-y)m_B^2-m_{q}^2+i\epsilon)},\nonumber\\[0.2cm]
A_{SP}^{P_1 V_2}(M) \sim \int dxdy \frac{(\mu_{P_1}-3(2x-1)m_{V_2})y(1-y)} {(x(1-y)m_B^2-\mu_g^2+i\epsilon)((1-y)m_B^2-m_{q}^2+i\epsilon)},\nonumber\\[0.2cm]
A_{SP}^{V_1 P_2}(M) \sim \int dxdy \frac{(-3(1-2x)m_{V_1}-\mu_{P_2})y(1-y)} {(x(1-y)m_B^2-\mu_g^2+i\epsilon)((1-y)m_B^2-m_{q}^2+i\epsilon)},\nonumber\\[0.2cm]
A_{SP}^{V_1 V_2}(M) \sim \int dxdy \frac{3(1-2x)(-m_{V_1}+3(2x-1)m_{V_2})y(1-y)} {(x(1-y)m_B^2-\mu_g^2+i\epsilon)((1-y)m_B^2-m_{q}^2+i\epsilon)}.
\end{eqnarray}
Since the contributions of these amplitudes are dominated by the area $x\sim0$ or $y\sim1$, $A_{SP}^{P_1 P_2}(M)$ and $A_{SP}^{P_1 V_2}(M)$ have the same sign, while $A_{SP}^{V_1 P_2}(M)$ and $A_{SP}^{V_1 V_2}(M)$ have a different sign from $A_{SP}^{P_1 P_2}(M)$. As a result, we use the same strong phase for $A_{SP}^{P_1 P_2}(M)$ and $A_{SP}^{P_1 V_2}(M)$, and another one for $A_{SP}^{V_1 P_2}(M)$ and $A_{SP}^{V_1 V_2}(M)$.

\section{Theoretical Input Parameters} \label{Input}

The numerical predictions in our calculations depend on a set of input parameters, such as the Wilson coefficients, the CKM matrix elements, the hadronic parameters, and so on. Here we present all
the relevant input parameters as follows.

\subsection{Light-Cone Distribution Amplitudes}

For the $B_s$-meson wave function, we take the following standard form in our numerical calculations~\cite{h.y.cheng}:
\begin{equation}
\phi_{B_s}(x)=N_Bx^2(1-x)^2
\exp\left[-\frac{1}{2}\left(\frac{xm_{B_s}}{\omega_{B_s}}\right)^2 \right]
\;, \label{def:lcda1}
\end{equation}
where the shape parameter $\omega_{B_s}=0.45$~GeV, and $N_B$ is a normalization constant.

We next specify the light-cone distribution amplitudes~(LCDAs) for pseudoscalar and vector mesons. The general expressions of twist-2 LCDAs are
\begin{eqnarray}
\phi_{P}(x,\mu) &=& 6x(1-x) \big[1 + \sum^{\infty}_{n=1} a^P_n(\mu) C^{3/2}_n(2x-1)\big],\nonumber\\
\phi_{V}(x,\mu) &=& 6x(1-x) \big[1 + \sum^{\infty}_{n=1} a^V_n(\mu) C^{3/2}_n(2x-1)\big],\nonumber\\
\phi_{V}^T(x,\mu) &=& 6x(1-x) \big[1 + \sum^{\infty}_{n=1} a^{T,V}_n(\mu) C^{3/2}_n(2x-1)\big],
\end{eqnarray}
and those of twist-3 ones are
\begin{eqnarray}
\phi_{p}(x,\mu) &=&1,\  \hspace{0.5cm}
\phi_{\sigma}(x,\mu) = 6x(1-x),\nonumber\\
\phi_{\nu}(x,\mu) &=& 3 \big[2x - 1 + \sum^{\infty}_{n=1} a^{T,V}_n(\mu) P_{n+1}(2x-1)\big]\nonumber\\
\phi_{+}(x)& =& 3 (1-x)^2, \hspace{0.5cm} \phi_{-}(x) = 3 x^2,
\end{eqnarray}
where $C_n$(x) and $P_n$(x) are the Gegenbauer and Legendre polynomials, respectively. The shape parameters of light mesons are taken from~\cite{Ball:2007rt} and listed in Table~\ref{parameters}.

\begin{table}[t]
\begin{center}
\caption{\small{Values of Gegenbauer moments at the scale $\mu$=1 GeV taken from \cite{Ball:2007rt} and at $\mu$=1.5GeV via a running}}\label{parameters}
\begin{tabular}{l|c|c|c|c|c|c|c}
\hline \hline
      &$\mu$ & $\pi$ &K &$\rho$&$K^*$&$\phi$ &$\omega$  \\
%\hline $\mu$ & 1GeV &1.5GeV& 1GeV &1.5GeV& 1GeV &1.5GeV& 1GeV &1.5GeV& 1GeV &1.5GeV& 1GeV &1.5GeV \\
\hline
$a_1$&1.0 &  --  &$0.06\pm0.03$&--&$0.03\pm0.02$&--&-- \\
     \cline{2-2} &1.5& --&$0.05\pm0.03$ &--&$0.03\pm0.02$&--& --\\
\hline
$a_2$&1.0&$0.25\pm0.15$&$0.25\pm0.15$&$0.15\pm0.07$&$0.11\pm0.09$&$0.15\pm0.07$&$0.18\pm0.08$  \\
      \cline{2-2} &1.5&$0.20\pm0.12$&$0.20\pm0.12$&$0.12\pm0.05$&$0.09\pm0.07$&$0.12\pm0.05$&$0.14\pm0.06$\\
\hline
$a_1^T$&1.0  &--&--&-- &$0.04\pm0.03$&--&-- \\
      \cline{2-2} &1.5& --&--& --&$0.03\pm0.03$&--&--\\
\hline
$a_2^T$&1.0&--&-- &$0.14\pm0.06$&$0.10\pm0.08$&$0.16\pm0.06$&$0.14\pm0.07$ \\
     \cline{2-2} &1.5&--& --&$0.11\pm0.05$&$0.08\pm0.06$&$0.13\pm0.05$&$0.11\pm0.05$\\
\hline \hline
\end{tabular}
\end{center}
\end{table}

The parameters in Table~\ref{parameters} are given at the scale $\mu$=1.0~GeV and $\mu$=1.5~GeV, where the values at $\mu$=1.0~GeV are taken from ref.\cite{Ball:2007rt}, which should run to the physical scale in the B meson decays with $\mu \simeq \sqrt{2\Lambda_{QCD}m_b}$. In our numerical calculations, we take $\mu =1.5\pm 0.1{\rm GeV}$ which is corresponding to $\Lambda_{QCD}\simeq 288^{+21}_{-18}{\rm MeV}$ evaluated from the data $\alpha_s(M_z) = 0.1172\pm 0.002$. It is noted that LCDAs of light mesons become much closer to their asymptotic forms~(all shape parameters become zero) when the scale runs to higher values.

\subsection{Decay Constants and Other Input Parameters}

For decay constants of various mesons and other hadronic parameters, we list them in Table~\ref{input}. As for the CKM matrix elements, we shall use the Wolfenstein parametrization~\cite{Wolfenstein:1983yz} with the values of four parameters\cite{Charles:2004jd}:
$A=0.798^{+0.023}_{-0.017}$, $\lambda=0.2252^{+0.00083}_{-0.00082}$, $\bar
\rho=0.141^{+0.035}_{-0.021}$, and $\bar \eta=0.340\pm0.016$.

%%%%%%%%%%%%%%%%%%%%%%%%%%%%%%%%%%%%%%%%%%
\begin{table}[ht]
\begin{center}
\caption{The hadronic input parameters~\cite{Amsler:2008zzb} and the decay constants taken from the QCD sum rules~\cite{Ball:2006eu,ball3} and Lattice theory~\cite{lattice}.} \label{input}
\doublerulesep 0.8pt
\tabcolsep 0.08in \vspace{0.2cm}
\begin{tabular}{ccccccc} \hline\hline
$\tau_{B_s}$ &$m_{B_s}$  &$m_b$ &$m_t$&$m_u$ &$m_d$&$m_c$\\ \hline

$1.472$ps &$5.366$GeV  &$4.4$GeV &$173.3$GeV&$4.2$MeV &$7.6$MeV &$1.5$GeV\\

$m_s$& $m_{\pi^\pm}$ & $m_{\pi^0}$ & $m_K $&$m_{\rho^0}$ &$m_{\rho^{\pm}}$&$m_{\omega}$ \\

$0.122$GeV& $0.140$GeV & $0.135$GeV &$0.494$GeV &$0.775$GeV &$0.775$GeV &1.7GeV\\

$m_{\phi}$&$m_{K^{*\pm}}$&$m_{K^{*0}}$& $\mu_\pi$ &$\mu_K$& $f_{B_s}$ &$f_{\pi}$\\

1.8GeV &$300$MeV &$0.78$GeV &$1.02$GeV&$0.892$GeV&$0.230$GeV & $0.130$GeV \\

$f_K$ &$f_{\rho}$&$f_{\omega}$&$f_{K^*}$&$f_{\phi}$ &$f_{K^*}^T$&$f_{\phi}^T$\\

$0.16$GeV&$0.216$GeV&$0.187$GeV&$0.220$GeV&$0.215$GeV &$0.185$GeV &$0.186$GeV \\

$f_{\rho}^T$ &$f_{\omega}^T$\\

$0.165$GeV&$0.151$GeV \\   \hline\hline
\end{tabular}
\end{center}
\end{table}
%%%%%%%%%%%%%%%%%

In our numerical calculations, the running scale is taken to be
\begin{eqnarray}
\mu = 1.5\pm0.1{\rm GeV} \sim \sqrt{2\Lambda_{QCD}m_b}.
\end{eqnarray}
The scale of $\alpha_{s}(\mu)$ in the six-quark operator effective
Hamiltonian is also taken at $\mu = 1.5$ GeV. The mass of b quark used here is the running mass at $\mu=1.5$~GeV and evaluated following the framework in \cite{quarkmass} as:
\begin{eqnarray}
m^{}_q(\mu) &=& {{\cal R}(\alpha^{}_s(\mu))} \hat{m}^{}_q \
,\nonumber\\
 {\cal
R}(\alpha^{}_s) &=&
\left(\frac{\alpha^{}_s}{\pi}\right)^{\gamma^{}_0/\beta^{}_0}
\left[1 + \frac{\alpha^{}_s}{\pi}{\cal C}^{}_1   \right] \; .
%+\frac{\alpha^2_s}{2\pi^2} \left({\cal C}^2_1 + {\cal C}^{}_2\right)
%+ \frac{\alpha^3_s}{\pi^3} \left( \frac{1}{6} {\cal C}^3_1 +
%\frac{1}{2} {\cal C}^{}_1 {\cal C}^{}_2 + \frac{1}{3}{\cal C}^{}_3
%\right) \right] \; .
\end{eqnarray}
The definition of $C_1$ can be found in \cite{quarkmass}. Numerically,
we find that $m_{b}(\mu)\simeq$5.44 GeV at NLO when $\mu =1.5$ GeV.

In addition, the infrared cut-offs for gluon and light quarks are
the basic scale to determine annihilation diagram contributions~(the
smaller $\mu_g$, the larger contributions). From our previous work~\cite{Su:2010vt}, it is noted that reasonable predictions can be made when taking $\mu_q=\mu_g$=0.37 GeV. Here we shall use the same values for $\mu_q$ and $\mu_g$.

\subsection{Form Factors}

As it is known that the transition form factors for the so-called factorizable contributions in the usual four quark effective Hamiltonian approach have to be provided from outside in QCDF
and SCET (such as by resorting to QCD sum rules or
lattice QCD). The method developed based on the six quark effective Hamiltonian allows us to calculate the relevant transition form
factors via a simple factorization approach. They are calculated by the following formalisms
 \begin{eqnarray}
&&F_0^{B_s\rightarrow M_1}=\frac{4\pi\alpha_s(\mu)C_F}{N_c m_{B_s}^2
 F_{M_2}}T_{LL}^{FM_1M_2}(B_s)(M_1,M_2 = P),\nonumber\\
&&V^{B_s\rightarrow M_1}=\frac{4\pi\alpha_s(\mu)C_F}{N_c m_{B_s}^2
F_{M_2}}T_{LL,\perp}^{FM_1M_2}(B_s)\frac{m_{B_s}^2(m_{B_s}+m_{M_1})}{m_{M_2}(m_{B_s}^2-m_{M_1}m_{M_2})}(M_1,M_2 = V),\nonumber\\
&&A_0^{B_s\rightarrow M_1}=\frac{4\pi\alpha_s(\mu)C_F}{N_c m_{B_s}^2
F_{M_2}}T_{LL}^{FM_1M_2}(B_s)(M_1 = V,M_2 = P),\nonumber\\
&&A_1^{B_s\rightarrow M_1}=\frac{4\pi\alpha_s(\mu)C_F}{N_c m_{B_s}^2
F_{M_2}}T_{LL,//}^{FM_1M_2}(B_s)\frac{m_{B_s}^2}{m_{M_2}(m_{B_s}+m_{M_1})}(M_1,M_2
= V),
\end{eqnarray}
with:
 \begin{eqnarray}
  T_{LL,\perp}=\frac{1}{2}(T_{LL,+}-T_{LL,-}), && C_F=\frac{N_c^2-1}{2
  N_c}, \nonumber
\end{eqnarray}
where the amplitudes $T_{LL}^{FM_1M_2}$ are given in the appendix of ~\cite{Su:2010vt}(see eq. (A36)). Before giving predictions of the observables in $B_s \to PP, PV, VV$ decays, we first present our numerical results for the form factor at
$q^2=0$ in Table~\ref{tab:formfactor}. For a comparison, we also list the results calculated from the QCD sum-rules, light-cone sum rules~\cite{Ball:2004rg,formfac1,Wu:2006rd}. In ref.\cite{Wu:2006rd}, the heavy bottom quark was treated based on the heavy quark effective field theory (HQEFT) via $1/m_Q$ expansion. Within their respective uncertainties, the results obtained in our present approach are consistent with the ones from all other approaches.

%%%%%%%%%%%%%%%%%%%%%%%%%%%%%%%%
\begin{table}
\vspace{-2em}
\begin{center}
\caption{\small{The $B_s \to P,V$ form factors at $q^2 = 0$ in QCD Sum
Rules, Light Cone and present work, where the errors stem mainly from the
uncertainties in the global parameters $\mu_{scale}=1.5\pm0.1
\text{GeV}, \mu_g=0.37\pm0.037\text{GeV}$, and the shape
parameters of light mesons. Within their respective uncertainties,
our predictions are consistent with the results from the other approaches. }}\label{tab:formfactor} \vspace{0.2cm}
%\scriptsize{ \vspace{0.2cm}
\begin{tabular}{c|c|c|c|c|c|c}
\hline \hline
\ \ \ \ \ Mode          & F(0)  & QCDSR\cite{Ball:2004rg}&LC\cite{formfac1}& LC\cite{Wu:2006rd} & This work  \\
\hline
 $B_s \to K^*$ & V&0.311&0.323&0.285 &$0.227^{+0.064+0.003}_{-0.037-0.002}$\\
  \cline{2-6}&$A_0$&0.360&0.279&0.222 &$0.280^{+0.082+0.013}_{-0.043-0.008}$\\
  \cline{2-6}&$A_1$&0.233&0.228&0.227 &$0.178^{+0.046+0.002}_{-0.027-0.002}$\\
\hline
 $B_s \to \phi$ &V&0.434&0.329&0.339 &$0.259^{+0.080+0.006}_{-0.036-0.003}$\\
  \cline{2-6}&$A_0$&0.474&0.279&0.212 &$0.311^{+0.096+0.014}_{-0.047-0.006}$\\
  \cline{2-6}&$A_1$&0.311&0.232&0.271 &$0.194^{+0.052+0.004}_{-0.028-0.002}$\\
  \hline
  $B_s \to K$ &$F_0$&$ $&0.290&0.296 &$0.260^{+0.053+0.007}_{-0.031-0.003}$\\
  \hline \hline
\end{tabular}
%}
\end{center}
\end{table}
%%%%%%%%%%%%%%%%%

\section{Numerical Results and Discussions}\label{sec:fitting}

In this section, we shall classify the 24 channels of $B_s$ decays into two light mesons according to the final states, and give our predictions for the branching ratios, the CP asymmetries, and the longitudinal polarization fractions. Since there are only a few data available for $B_s$ decays, we shall make comparisons with other theoretical predictions. The comparisons with the current experimental data, if possible, are also made. In addition, we shall discuss the SU(3) flavour symmetry in the decays which have been experimentally observed and also in some other interesting ones. According to different decay modes, we shall give our predictions for the observables one by one.

\subsection{$B_s\to PP$ decays}

This type of decays has been discussed in previous paper~\cite{Su:2008mc}. In this work, we will reexamine them and shed light on the influences of effective Wilson coefficients and annihilation contribution with a strong phase. The resulting branching ratios and CP asymmetries of $B_s\to PP$ decays are listed in Table~\ref{tab:br41-1}. In order to have better test on our theoretical framework, we also list the most recent predictions based on QCDF with strong phase effects~\cite{Cheng:2009mu} and the predictions from pQCD~\cite{Ali:2007ff} approach in Table~\ref{tab:br41}. The first theoretical error in our calculations is referred to the global parameters of running energy scale $\mu_{scale}$ and the infrared energy scale $\mu_g$, and the second one is from the shape parameters of light mesons.

%%%%%%%%%%%%%%%%%%%%%%%%%%%%%%%
\begin{table}
\begin{center}
\vspace{-1em} \caption{\small {The branching ratio~(in units of $10^{-6}$) and direct CP asymmetries~(in $\%$) in $B_s\to PP$ decays. The central values are obtained at $\mu_q=\mu_g$=0.37GeV, the first error stems from the uncertainties in the global parameters $\mu_{scale}=1.5\pm0.1 \text{GeV}, \mu_g=0.37\pm0.037\text{GeV}$, and the second from the shape parameters of light mesons. ${\rm NLO}^{eff}$ and ${\rm NLO}^{eff}(\theta^a)$ stand for results with ``NLO correction+effective Wilson coefficients" and ``NLO correction+effective Wilson coefficients+annihilation with strong phase", respectively.}}\label{tab:br41-1} \scriptsize{
\begin{tabular}{l|c|c|c|ccc}
\hline \hline
\ \ \ \ \ Mode          & Data\cite{Morello:2006pv,dtonelli} &\multicolumn{5}{c}{This work} \\
\cline{3-7}
                        &                  &   NLO        &   NLO$^{eff}$   &   NLO$^{eff}(-10^\circ)$     & NLO$^{eff}(5^\circ)$  &NLO$^{eff}(20^\circ)$    \\
\hline

$B_s \to \pi^+ K^-$& $5.0\pm1.25$  & $7.7$  &7.0&7.0& $7.1$&7.2 \\
$B_s \to \pi^0 \bar{K}^0$& \textbf{-}    & $0.2$  &1.1&1.1& $1.1$&1.1\\
$B_s\to K^+K^-  $ & $24.4\pm1.4\pm3.5$  &  $20.8$&20.5&17.5&  $22.0$&26.2\\
$B_s \to K^0\bar{K}^0  $ & \textbf{-}    &  $22.6$ &20.7&17.5&  $22.3$&26.9\\
\hline

$A_{CP}(\pi^+K^-)$& $39\pm 17$ &  $24.3$ &28.5&29.3&  $27.8$ &24.7 \\
$A_{CP}(\pi^0\bar{K}^0)$& \textbf{-}     &  $77.6$  &61.6&56.0&  $64.4$&72.1\\
$A_{CP}(K^+K^-  )$& \textbf{-}    &$-14.4$    &-15.4&-18.2&$-14.2$ &-10.8 \\
$A_{CP}(K^0\bar{K}^0  )$& \textbf{-}    &$0$     &0&0&$0$&0\\
\hline \hline
\end{tabular}
}
\end{center}
\end{table}
%%%%%%%%%%%%%%%%%%%%%%
\begin{table}
\begin{center}
\tabcolsep 0.02in\vspace{-1em} \caption{\small{Comparisons of predictions between our framework and other methods in $B_s \to PP$ decays.}}\label{tab:br41} \scriptsize{
\begin{tabular}{l|c|c|c|c|ccc}
\hline \hline
\ \ \ \ \ Mode          & Data\cite{Morello:2006pv,dtonelli} &QCDF\cite{Cheng:2009mu}&pQCD\cite{Ali:2007ff} &SCET\cite{Williamson:2006hb} &\multicolumn{3}{c}{This work} \\
\cline{6-8}
                        &                  &           &      &  & LO        & NLO &NLO($a^{eff},\theta^{a}$)    \\
\hline

$B_s \to \pi^+ K^-$& $5.0\pm1.25$  &$5.3^{+0.4+0.4}_{-0.8-0.5}$       &$7.6^{+3.2+0.7+0.5}_{-2.3-0.7-0.5}$ &$4.9\pm1.2\pm1.3\pm0.3$ &7.2       & $7.7$  & $7.1^{+3.2+0.6}_{-1.8-0.2}$ \\
$B_s \to \pi^0 \bar{K}^0$& \textbf{-}    &$1.7^{+2.5+1.2}_{-0.8-0.5}$ &$0.16^{+0.05+0.10+0.02}_{-0.04-0.05-0.01}$ & $0.76\pm0.26\pm0.27\pm0.17$ &0.2       & $0.2$  & $1.1^{+0.3+0.6}_{-0.2-0.1}$\\
$B_s\to K^+K^-  $ & $24.4\pm1.4\pm3.5$  &$25.2^{+12.7+12.5}_{-7.2-9.1}$       &$13.6^{+4.2+7.5+0.7}_{-3.2-4.1-0.2}$ &$18.2\pm6.7\pm1.1\pm0.5$ &16.6     &  $20.8$&  $22.0^{+5.6+11.8}_{-3.4-3.0}$\\
$B_s \to K^0\bar{K}^0  $ & \textbf{-}    &$26.1^{+13.5+12.9}_{-8.1-9.4}$       &$15.6^{+5.0+8.3+0.0}_{-3.8-4.7-0.0}$ &$17.7\pm6.6\pm0.5\pm0.6$ &18.2    &  $22.6$ &  $22.3^{+4.3+12.2}_{-5.3-3.2}$\\
$B_s \to \pi^+ \pi^-$    & $0.5\pm 0.5$  &$0.26^{+0.00+0.10}_{-0.00-0.09}$ &$0.57^{+0.16+0.09+0.01}_{-0.13-0.10-0.00}$&$$ &0.18       & $0.23$  & $0.23^{+0.01+0.26}_{-0.01-0.011}$\\
$B_s \to \pi^0 \pi^0$    & \textbf{-}    &$0.13^{+0.0+0.05}_{-0.0-0.05}$ &$0.28^{+0.08+0.04+0.01}_{-0.07-0.05-0.00}$ &$$ &0.09      & $0.12$ & $0.12^{+0.01+0.13}_{-0.01-0.05}$\\
\hline

$A_{CP}(\pi^+K^-)$& $39\pm 17$ &$20.7^{+5.0+3.9}_{-3.0-8.8}$&$24.1^{+3.9+3.3+2.3}_{-3.6-3.0-1.2}$ &$20\pm17\pm19\pm5$ &21.5       &  $24.3$ &  $27.8^{+6.0+6.9}_{-2.0-4.1}$  \\
$A_{CP}(\pi^0\bar{K}^0)$& \textbf{-}     &$36.3^{+17.4+26.6}_{-18.2-24.3}$&$59.4^{+1.8+7.4+2.2}_{-4.0-1.3-3.5}$ &$-58\pm39\pm39\pm13$ &1.2       &$77.6$  &  $64.4^{+2.0+6.0}_{-1.8-11.6}$\\
$A_{CP}(K^+K^-  )$& \textbf{-}    &$-7.7^{+1.6+4.0}_{-1.2-5.1}$&$-23.3^{+0.9+4.9+0.8}_{-0.2-4.4-1.1}$&$-6\pm5\pm6\pm2$ &-15.7           &$-14.4$    &$-14.2^{+0.1+1.7}_{-0.1-0.4}$  \\
$A_{CP}(K^0\bar{K}^0  )$& \textbf{-}    &$0.4^{+0.04+0.10}_{-0.04-0.04}$&0 &$<10$                                                &0.0           &$0.0$     &$0.0^{+0+0}_{-0-0}$\\
$A_{CP}(\pi^+\pi^-    )$& \textbf{-}     &   $0$     &$-1.2^{+0.1+1.2+0.1}_{-0.4-1.2-0.1}$& &5.2   & $4.5$  & $4.5^{+0.4+1.5}_{-0.2-0.5}$  \\
$A_{CP}(\pi^0\pi^0    )$& \textbf{-}     &   $0$  &$-1.2^{+0.1+1.2+0.1}_{-0.4-1.2-0.1}$ & &5.2    & $4.5$   & $4.5^{+0.4+1.5}_{-0.2-0.5}$ \\
\hline \hline
\end{tabular}
}
\end{center}
\end{table}

%%%%%%%%%%%%%%%%%%%%%%

\subsubsection*{1. $B_s \to K^- \pi^+, \bar K^0\pi^0$}

The tree-dominated $B_s \to K^-\pi^+$ decay has sizable branching ratios of order $7.6\times 10^{-6}$ at LO in our framework, which is above the experimental result $(5.0\pm1.25)\times 10^{-6}$. When considering the NLO contribution, one can see that the contribution would further worsen the prediction. After including the strong phase effect which cancels out the NLO contribution, then the prediction is a little smaller than the LO one while still bigger than the data. However, our result is in good agreement with the ones obtained from the other methods, such as QCDF~\cite{Beneke:2003zv} and pQCD~\cite{Ali:2007ff}. In addition, the contribution of penguin operators is comparable to the one of tree operators, and hence the interference between the two contributions is large; as a result, a big direct CP asymmetry is predicted in this decay mode. Moreover, the annihilation contribution with a strong phase has insignificant effect on the branching ratios and CP asymmetries.

As it is well-known that the decay $B_s \to K^-\pi^+$ can be related to $B_d\to \pi^-\pi^+$ by SU(3) symmetry which implies $A(B_s\to K^-\pi^+)\approx A(B_d\to \pi^-\pi^+)$~\cite{Gronau:2000zy,He:2000dg}. The relation between the two amplitudes result in
\begin{eqnarray}
Br(B_s\to K^-\pi^+)\approx Br(B_d\to \pi^-\pi^+), \hspace{0.5cm}
A_{CP}(B_s\to K^-\pi^+)\approx A_{CP}(B_d\to \pi^-\pi^+).
\end{eqnarray}
These relations are satisfied experimentally. From the Table~\ref{tab:br41} and our previous paper~\cite{Su:2010vt}, we have $Br(B_d\to \pi^-\pi^+) = 6.6^{+3.3}_{-1.3}$, $A(B_d\to \pi^-\pi^+) = 26.0^{+4.3}_{-3.2}$ for branching ratios and direct CP asymmetries, respectively. So the ratios:
\begin{eqnarray}
R_{Br} =  \frac{Br(B_s\to K^-\pi^+)}{Br(B_d\to \pi^-\pi^+)} \approx 1.08,\hspace{0.5cm}
R_{A_{CP}} =  \frac{A_{CP}(B_s\to K^-\pi^+)}{A_{CP}(B_d\to \pi^-\pi^+)} \approx 1.07,
\end{eqnarray}
indicates that the SU(3) symmetry relations are satisfactorily respected in our framework.

For $B_s \to \bar K^0\pi^0$, its branching ratio is much smaller than the ones of $B_s \to K^-\pi^+$, since it is a color-suppressed decay mode. The contributions of effective Wilson coefficients and annihilation with a strong phase enhance the rate by a factor of about 5.5. As for the direct CP asymmetry, the NLO contribution, as well as the effective Wilson coefficients and annihilation with a strong phase have remarkable effects on it. It is interesting to note that the prediction of branching ratio in our framework is roughly consistent with the one in SCET~\cite{Williamson:2006hb}, while the results of direct CP asymmetry have opposite signs in these two methods.

\subsubsection*{2. $B_s \to K^- K^+, \bar K^0 K^0$}

These two decay channels are penguin-dominated, and their branching ratios are of order $23\times 10^{-6}$ in our framework after including the effective Wilson coefficients and annihilation contribution with a strong phase. Such large branching rates can be easily measured at future LHC-b and SuperB experiments. Moreover, the annihilation contributions with a strong phase have remarkable effects on the branching ratios in these two decay modes. In our method, the prediction of direct CP asymmetry in $B_s\to K^+K^-$ decay has the same sign as the ones in other methods, and the one in $B_s\to \bar K^0 K^0$ decay vanishes because of the absence of interference between the tree and penguin amplitudes. It has been argued that the CP asymmetry of the decay $B_s\to \bar K^0 K^0$ is a very promising observable to look for effects of new physics~\cite{Baek:2006pb,Ciuchini:2007hx}. For example, it is shown in~\cite{Baek:2006pb} that the direct CP violation of $B_s\to \bar K^0K^0$, which is not more than $1\%$ in the SM, can be 10 times larger in the presence of SUSY contributions while its rate remains unaffected.

In addition, the decay $B_s\to K^+K^-$ can be related to $B_d\to \pi^- K^+$ by SU(3) symmetry, which implies that
\begin{eqnarray}
Br(B_s\to K^-K^+)\approx Br(B_d\to \pi^-K^+), \hspace{0.5cm}
A_{CP}(B_s\to K^-K^+)\approx A_{CP}(B_d\to \pi^-K^+),
\end{eqnarray}
the first relation is experimentally satisfied $24.4\pm1.4\pm3.5\approx19.4\pm0.6$. Our predictions~\cite{Su:2010vt} for the the two ratios are:
\begin{eqnarray}
R_{Br} =  \frac{Br(B_s\to K^-K^+)}{Br(B_d\to \pi^-K^+)} \approx 1.07,\hspace{0.5cm}
R_{A_{CP}} =  \frac{A_{CP}(B_s\to K^-K^+)}{A_{CP}(B_d\to \pi^-K^+)} \approx 1.08,
\end{eqnarray}
indicates that the SU(3) symmetry relations are satisfactorily respected in our framework.

%for the two relations are $22.0^{+5.6}_{-3.4}\approx20.5^{+5.2}_{-3.0}$ and $-14.2^{+0.1}_{-0.1}\approx-13.1^{+0.9}_{-0.3}$, which are both in good agreement with the SU(3) symmetry.

\subsubsection*{3. $B_s\to \pi^+\pi^-, \pi^0\pi^0$}

The $B_s\to \pi\pi$ decays are pure annihilation processes. The predictions of $Br(B_s\to \pi^+\pi^-)=2.8\times 10^{-7}$ and $Br(B_s\to \pi^0\pi^0)=1.4\times 10^{-7}$ are in good agreement with the QCDF results~\cite{Cheng:2009mu}. Moreover, the direct CP asymmetry are small in these decays. Being of pure annihilation processes, it is meaningless to discuss the phase influence in these decays, we thus do not list them in Table~\ref{tab:br41-1}.

\subsection{$B_s \to PV$ decays}

We now turn to discuss the observables in $B_s\to PV$ decays. Since the final states are a pseudoscalar and a vector meson, only the longitudinal polarization of the vector meson can contribute. In general, there are two kinds of emission diagrams as the Lorentz structure of the vector meson LCDAs is different from the pseudoscalar case. If the emitted meson is a vector meson, the effective Wilson coefficient is the same as that of $B_s\to PP$ case because they are both characterized by the $B_s\to P$ transition form factors. However, when the emitted meson is a pseudoscalar meson, which is characterized by the $B_s\to V$ transition form factors, the effective Wilson coefficient is smaller than that of $B_s\to PP$ in our framework, the detail can be found in Eq.~(\ref{aeff}). The predictions for the branching ratios and direct CP asymmetries in $B_s\to PV$ decays are listed in Tables~\ref{tab:br42-1}-\ref{tab:br43}.

%%%%%%%%%%%%%%%%%%%%%%%%%%%%
\begin{table}
\caption{\small{The branching ratios~(in units of $10^{-6}$) and direct CP asymmetries~($\%$) in $B_s\to \pi K^*, \rho K$ decays. We use different strong phase in these decays since they are characterized by the $B_s\to V$ and $B_s\to P$ transition form factors, respectively. The other captions are the same as Table \ref{tab:br41-1}.} }\label{tab:br42-1}
\begin{center}
\scriptsize{
 \begin{tabular}{c|c|c|ccc}
 \hline\hline {Mode}   &\multicolumn{5}{c}{This work} \\
\cline{2-6}
                                         &   NLO        &   NLO$^{eff}$   &   NLO$^{eff}(45^\circ)$     & NLO$^{eff}(60^\circ)$ &NLO$^{eff}(75^\circ)$   \\
\hline
   $B_s\to\pi^+K^{*-}$             &  8.2&7.3&7.2&$7.2$&7.3\\
   $B_s\to{\pi}^{0}\bar K^{*0}$         & 0.2&0.3&0.3&$0.3$&0.3\\
   $A_{CP}(\pi^-K^{*+})$                  & -27.5&-28.3&-18.7&$-12.8$&-6.0\\
   $A_{CP}({\pi}^{0}\bar K^{*0})$           & -29.3&66.5&15.4&$-3.8$&-22.7\\
 \hline
                                         &   NLO        &   NLO$^{eff}$   &
                      NLO$^{eff}(-10^\circ)$     & NLO$^{eff}(5^\circ)$ &NLO$^{eff}(20^\circ)$  \\
\hline
   $B_s\to\rho^+K^{-}$              & 19.7&17.5&17.5&$17.6$&17.7\\
   $B_s\to{\rho}^{0}\bar K^{0}$          & 0.4&0.6&0.5&$0.6$&0.6\\
   $A_{CP}(\rho^-K^{+})$                & 17.9&18.9&19.2&$18.5$&16.4  \\
   $A_{CP}({\rho}^{0}\bar{K}^{0})$           & 77.0&-29.7&-36.8&$-25.8$&-12.9\\
 \hline\hline
\end{tabular}}
\end{center}
\end{table}
%%%%%%%%%%%%%%%%%%%%%%%
\begin{table}
\caption{\small{The same as Table \ref{tab:br42-1} but for $B_s\to K K^*, \rho \pi$ decays. Here we use ($5^\circ$, $60^\circ$) as the default inputs.}} \label{tab:br42-2}
\begin{center}\scriptsize{
 \begin{tabular}{c|c|c|ccccc}
 \hline\hline {Mode}   &\multicolumn{7}{c}{This work} \\
\cline{2-8}
                        &    NLO         &   NLO$^{eff}$       &   ($5^\circ$, $60^\circ$)   &   ($-10^\circ$, $60^\circ$)&   ($20^\circ$, $60^\circ$)&   ($5^\circ$, $45^\circ$)&   ($5^\circ$, $75^\circ$)   \\
\hline
   $B_s\to K^{*-} K^+$                       & 5.8&6.6&$7.8$&7.8&7.8&7.7&7.8\\
   $B_s\to K^- K^{*+}$                       & 8.1&7.6&$8.2$&6.5&9.7&8.2&8.2\\
   $B_s\to K^{*0}\bar{K}^{0}$           & 9.0&7.9&$8.5$&6.7&10.3&8.5&8.5\\
   $B_s\to K^{0}\bar{K}^{*0}$             & 5.6&7.5&$7.1$&7.1&7.1&7.4&6.6\\
   $B_s\to\rho^-\pi^+$                   & 0.04&0.04&$0.008$&0.006&0.014&0.014&0.006\\
   $B_s\to\pi^-\rho^+$                    & 0.04&0.04&$0.006$&0.003&0.011&0.011&0.003\\
   $B_s\to\pi^0\rho^0$                   & 0.04&0.04&$0.006$&0.003&0.012&0.012&0.005\\
 \hline
   $A_{CP}(K^+ K^{*-})$                        & 54.0&48.8&$23.2 $ &23.1&23.2&31.4&14.0\\
   $A_{CP}(K^{*+} K^-)$                        & -32.6&-32.2&$-29.5$ &-38.2&-22.0&-29.5&-29.5\\

   $A_{CP}(K^{0}\bar{K}^{*0})$              & 0&0&$0$&0&0&0&0\\
   $A_{CP}(K^{*0}\bar{K}^0)$             & 0&0&$0$&0&0&0&0\\

   $A_{CP}(\rho^+\pi^-)$                       & -1.9&-1.9&$-0.6$&-0.2&-1.1&-1.1&-0.2\\
   $A_{CP}(\pi^+\rho^-)$                      & -1.6&-1.6&$-0.4$&-0.3&-0.7&-0.7&-0.3\\
   $A_{CP}(\pi^0\rho^0)$                        & -1.7&-1.7&$-0.6$&-0.3&-0.9&-0.9&-0.3\\
 \hline
 \hline
\end{tabular}}
\end{center}
\end{table}

%%%%%%%%%%%%%%%%%%%%%%%%%%%%
\begin{table}
\caption{\small{The branching ratios~(in units of $\times 10^{-6}$) of $B_s\to PV$ decays. For comparison, we also quote the theoretical estimates of the branching ratios in the
QCDF~\cite{Cheng:2009mu} and pQCD~\cite{BsPV} frameworks.} }\label{tab:br42}
\begin{center}\scriptsize{
 \begin{tabular}{c|c|c|ccc}
 \hline\hline {Mode}   &   QCDF   & pQCD   &\multicolumn{3}{c}{This work} \\
\cline{4-6}
                        &                             &       & LO        & NLO &NLO($a^{eff},\theta^{a}$)    \\
\hline
   $B_s\to\pi^+K^{*-}$             &  $7.8^{+0.4+0.5}_{-0.7-0.7}$&  $7.6^{+2.9+0.4+0.5}_{-2.2-0.5-0.3}$&7.7&8.2&$7.2^{+5.6+0.7}_{-2.2-0.5}$\\
   $B_s\to{\pi}^{0}\bar K^{*0}$         & $0.89^{+0.80+0.84}_{-0.34-0.35}$& $0.07^{+0.02+0.04+0.01}_{-0.01-0.02-0.01}$&0.09&0.2&$0.3^{+0.1+0.1}_{-0.1-0.1}$\\
   $B_s\to\rho^+K^{-}$              & $14.7^{+1.4+0.9}_{-1.9-1.3}$ & $17.8^{+7.7+1.3+1.1}_{-5.6-1.6-0.9}$&18.7&19.7&$17.6^{+8.2+0.1}_{-4.6-0.1}$\\
   $B_s\to{\rho}^{0}\bar K^{0}$          & $1.9^{+2.9+1.4}_{-0.9-0.6}$&  $0.08^{+0.02+0.07+0.01}_{-0.02-0.03-0.00}$&0.2&0.4&$0.6^{+0.2+0.1}_{-0.1-0.1}$\\

   $B_s\to K^{*-} K^+$                       & $11.3^{+7.0+8.1}_{-3.5-5.1}$& $4.7^{+1.1+2.5+0.0}_{-0.8-1.4-0.0}$&5.4&5.8&$7.8^{+0.3+1.5}_{-0.5-1.1}$\\
   $B_s\to K^- K^{*+}$                       & $10.3^{+3.0+4.8}_{-2.2-4.2}$& $6.0^{+1.7+1.7+0.7}_{-1.5-1.2-0.3}$&5.9&8.1&$8.2^{+1.3+2.1}_{-2.3-2.0}$\\

   $B_s\to K^{*0}\bar{K}^{0}$           & $10.5^{+3.4+5.1}_{-2.8-4.5}$ & $7.3^{+2.5+2.1+0.0}_{-1.7-1.3-0.0}$&6.5&9.0&$8.5^{+1.8+1.5}_{-2.1-1.6}$\\
   $B_s\to K^{0}\bar{K}^{*0}$             & $10.1^{+7.5+7.7}_{-3.6-4.8}$ &  $4.3^{+0.7+2.2+0.0}_{-0.7-1.4-0.0}$&4.6&5.6&$7.1^{+0.2+1.3}_{-0.4-1.1}$\\
   $B_s\to\rho^-\pi^+$                   & $0.02^{+0.00+0.01}_{-0.00-0.01}$&   $0.22^{+0.05+0.04+0.00}_{-0.05-0.06-0.01}$&0.03&0.04&$0.01^{+0.00+0.01}_{-0.00-0.00}$\\
   $B_s\to\pi^-\rho^+$                    & $0.02^{+0.00+0.01}_{-0.00-0.01}$ &  $0.24^{+0.05+0.05+0.00}_{-0.05-0.06-0.01}$&0.03&0.04&$0.01^{+0.00+0.01}_{-0.00-0.00}$\\
   $B_s\to\pi^0\rho^0$                   & $0.02^{+0.00+0.01}_{-0.00-0.01}$&  $0.23^{+0.05+0.05+0.00}_{-0.05-0.06-0.01}$&0.03&0.04&$0.01^{+0.00+0.01}_{-0.00-0.00}$\\
 \hline\hline
\end{tabular}}
\end{center}
\end{table}
%%%%%%%%%%%%%%%%%%%%%%%%
\begin{table}
\caption{The same as Table~\ref{tab:br42} but for the direct $CP$ asymmetries~(in \%) in the $B_s\to PV$  decays. }
 \label{tab:br43}
\begin{center}\scriptsize{
 \begin{tabular}{c|c|c|ccc}
  \hline\hline {Mode}    &   QCDF   & pQCD   & \multicolumn{3}{c}{This work} \\
\cline{4-6}
                        &                             &       & LO        & NLO &NLO($a^{eff},\theta^{a}$)    \\
\hline
   $A_{CP}(\pi^-K^{*+})$                  & $-24.0^{+1.2+7.7}_{-1.5-3.9}$& $-19.0^{+2.5+2.7+0.9}_{-2.6-3.4-1.4}$&-24.9&-27.5&$-12.8^{+7.0+4.9}_{-5.2-3.5}$\\
   $A_{CP}({\pi}^{0}K^{*0})$           & $-26.3^{+10.8+42.2}_{-10.9-36.7}$& $-47.1^{+7.4+35.5+2.9}_{-8.7-29.8-7.0}$ &40.1&-29.3&$-3.8^{+6.1+7.5}_{-6.7-7.4}$\\

   $A_{CP}(\rho^-K^{+})$                & $11.7^{+3.5+10.1}_{-2.1-11.6}$& $14.2^{+2.4+2.3+1.2}_{-2.2-1.6-0.7}$ &17.2&17.9&$18.5^{+3.0+2.9}_{-2.6-2.7}$  \\
   $A_{CP}({\rho}^{0}\bar{K}^{0})$           & $28.9^{+14.6+25.0}_{-14.5-23.7}$& $73.4^{+6.4+16.2+2.2}_{-11.7-47.8-3.9}$ &-22.4&77.0&$-25.8^{+4.1+4.5}_{-4.1-4.8}$\\

   $A_{CP}(K^+ K^{*-})$                        & $25.5^{+9.2+16.3}_{-8.8-11.3}$& $55.3^{+4.4+8.5+5.1}_{-4.9-9.8-2.5}$ &52.5&54.0&$23.2^{+1.4+2.7}_{-1.4-2.7} $ \\
   $A_{CP}(K^{*+} K^-)$                        & $-11.0^{+0.5+14.0}_{-0.4-18.8}$&  $-36.6^{+2.3+2.8+1.3}_{-2.3-3.5-1.2}$  &-40.0&-32.6&$-29.5^{+8.5+4.3}_{-8.5-4.9}$ \\

   $A_{CP}(K^{0}\bar{K}^{*0})$              & $0.10^{+0.08+0.05}_{-0.07-0.02}$&  $0$&0&0&$0^{+0+0}_{-0-0}$\\
   $A_{CP}(K^{*0}\bar{K}^0)$             & $0.49^{+0.08+0.09}_{-0.07-0.12}$&  $0$&0&0&$0^{+0+0}_{-0-0}$\\

   $A_{CP}(\rho^+\pi^-)$                       & $-11.1^{+0.7+13.9}_{-0.8-15.7}$ &  $-1.3^{+0.9+2.8+0.1}_{-0.4-3.5-0.2}$  &-1.8&-1.9&$-0.6^{+0.1+1.1}_{-0.1-2.0}$\\
   $A_{CP}(\pi^+\rho^-)$                      & $10.2^{+0.8+12.7}_{-0.7-12.8}$& $4.6^{+0.0+2.9+0.6}_{-0.6-3.5-0.3}$   &-1.5&-1.6&$-0.4^{+0.1+1.5}_{-0.1-0.9}$\\
   $A_{CP}(\pi^0\rho^0)$                        & $0$&  $1.7^{+0.2+2.8+0.2}_{-0.8-3.6-0.1}$&-1.6&-1.7&$-0.6^{+0.1+1.3}_{-0.1-1.2}$\\
 \hline\hline
\end{tabular}}
\end{center}
\end{table}

%%%%%%%%%%%%%%%%%%%%%%%%%%%%%%%%%%%

\subsubsection*{1. $B_s\to K^{*-}\pi^+, \bar K^{*0}\pi^0$}

As for $B_s\to K^{*-}\pi^+, \bar K^{*0}\pi^0$ decays, the former is tree-dominated and the later is color-suppressed. Here we use smaller effective Wilson coefficients since they are both characterized by the $B_s\to V$ transition form factors. Following the same consideration in our previous work for $B$ meson decays\cite{Su:2010vt}, we take into our calculations a strong phase $\theta=60^\circ$ as the default case for this type of decays. The annihilation contributions with a strong phase have insignificant effects on the branching ratios but remarkable effects on the direct CP asymmetries in these decays. The $B_s\to K^{*-} \pi^+ $ mode has the branching ratio of order $7.2\times 10^{-6}$ and big direct CP asymmetry, and the predictions in our framework are in good agreement with the ones in QCDF~\cite{Cheng:2009mu} and pQCD~\cite{Ali:2007ff}. As for the color-suppressed $B_s\to \bar K^{*0}\pi^0$ decay, our predictions are different from the ones of the other methods~(pQCD and QCDF), especially for the direct CP asymmetry.

\subsubsection*{2. $B_s\to \rho^+ K^{-}, \rho^0\bar K^{0}$}

These two decays, being characterized by the $B_s\to P$ transition form factors, are tree-dominated and color-suppressed modes, respectively. The annihilation contributions with a strong phase have remarkable effects on the direct CP asymmetries, especially in $B_s\to \rho^0\bar K^{0}$ decay. For $B_s\to \rho^+ K^{-}$ decay, our predictions are consistent with the ones in QCDF and pQCD. But for the color-suppressed mode, the predictions are different from each other among the current theoretical methods. With this situation, it is expected that the future more precise experimental data will give us an unambiguous answer.

\subsubsection*{3. $B_s\to K^{*-} K^+, K^- K^{*+}, K^{*0}\bar{K}^{0}, K^{0}\bar{K}^{*0}$}

Now we shed light on the $B_s\to K^*K$ decays. The considerations of influence from annihilation contributions with strong phases are complex since there are both $B_s\to P$ and $B_s\to V$ transitions, and the related predictions are listed in Table~\ref{tab:br42-2}. Although these four decays are all penguin-dominated, they have sizable branching ratios of order $8 \times 10^{-6}$ due to the fact that the related CKM elements ($V^*_{tb}V_{ts}\sim \lambda^2$) are relatively large in these decays. Moreover, it is interesting to note that our predictions are smaller than that of QCDF~\cite{Cheng:2009mu} while larger than that in pQCD~\cite{Ali:2007ff}. As for the direct CP asymmetries, it is large for the former two modes since there are strong interference between penguin and tree amplitudes, which are consistent with the ones in the QCDF and pQCD methods. However, for the latter two decays, the direct CP asymmetries vanish since there is only one type of the combination of CKM matrix elements, $V^*_{tb}V_{ts}$.

As we know that the pairs related by SU(3) symmetry are $B_d\to K^{*+}\pi^-$, $B_s\to K^{*+}K^-$, and $B_d\to \rho^- K^+$, $B_s\to K^{*-}K^+$. The exact symmetry implies that the amplitudes approximately equal in each pair, and then the branching ratios and direct CP asymmetries are approximately equal, namely
\begin{eqnarray}
&&Br(B_s\to K^{*+}K^-)\approx Br(B_d\to K^{*+}\pi^-),\ \quad A_{CP}(B_s\to K^{*+}K^-)\approx A_{CP}(B_d\to K^{*+}\pi^-),\nonumber\\
&&Br(B_s\to K^{*-}K^+)\approx Br(B_d\to \rho^- K^+),\ \quad A_{CP}(B_s\to K^{*-}K^+)\approx A_{CP}(B_d\to \rho^- K^+).
\end{eqnarray}
Our predictions for the ratios of the above observables are
\begin{eqnarray}
&&R_{Br} =  \frac{Br(B_s\to K^{*+}K^-)}{Br(B_d\to K^{*+}\pi^-)} \approx 0.92,\ \quad R_{A_{CP}} =  \frac{A_{CP}(B_s\to K^{*+}K^-)}{A_{CP}(B_d\to K^{*+}\pi^-)} \approx 0.9,\nonumber\\
&&R_{Br} =  \frac{Br(B_s\to K^{*-}K^+)}{Br(B_d\to \rho^- K^+)} \approx 1.06,\ \quad R_{A_{CP}} =  \frac{A_{CP}(B_s\to K^{*-}K^+)}{A_{CP}(B_d\to \rho^- K^+)} \approx 0.8.
%&&8.2^{+1.3}_{-2.3}\doteq 8.9^{+2.4}_{-1.4},\ \quad -29.5^{+8.5}_{-8.5}\doteq -32.7^{+2.0}_{-0.8}, \nonumber\\
%&&7.8^{+0.3}_{-0.5}\doteq 7.3^{+0.8}_{-0.5},\ \quad  23.2^{+1.4}_{-1.4}\doteq 29.0^{+2.1}_{-2.0}.
\end{eqnarray}
%which shows that the SU(3) relations are satisfactorily respected in our framework.

\subsubsection*{4. $B_s\to\rho^-\pi^+, \pi^-\rho^+, \pi^0\rho^0$}

These three decays proceed only through annihilation contributions. In each decay mode, there are both $B_s\to P$ and $B_s\to V$ transitions, which have different non-perturbative corrections as can be seen from Eq.~(\ref{aeff}). As a result, the influences from annihilation contributions with strong phase do not vanish even though these decays are pure annihilation processes, and the related discussions are also listed in Table~\ref{tab:br42-2}. The branching ratios are at the order of $10^{-8}$ and the direct CP asymmetries are small in these decays.

\subsection{$B_s\to VV$ Decays}

There are several other observables besides the branching ratios and CP asymmetries in $B_s\to VV$ decays, such as the polarization fractions and relative phases. Naive factorization without annihilation contribution predicts a longitudinal polarization fraction near 100\% for all $B_s\to VV$ decay modes, while the polarization anomaly in $B_s \to \phi \phi$~(the longitudinal polarization fraction $f_L$ is about 35\%) has been observed by the CDF~\cite{phiphip} experiments. Motivated by the anomaly, we shall study in detail the polarization, branching ratios and direct CP asymmetries in $B_s\to VV$ decays in this section. The effects of different strong phases on branching ratio, direct CP asymmetry and longitudinal polarization are listed in Table~\ref{tab:br44-1}. For there are no experimental data for the most of $B_s\to VV$ decays, the comparisons of predictions in different theoretical methods are especially important, which are listed in Table~\ref{tab:br44}.

%%%%%%%%%%%%%%%%%%%%%%%%%%%%%%%
\begin{table}[tb]
\caption{\small{CP-averaged branching ratios(in units of $\times 10^{-6}$), direct CP asymmetries(in $\%$) and the polarization fractions(in $\%$) for $B_s\to VV$ decays. The central values are obtained with $\mu_g^a$=0.52GeV and $\theta^{a} = 60^{\circ}$.}}\label{tab:br44-1}
 \begin{center}\scriptsize{
\begin{tabular}{c|c|c|c|ccc}
\hline \hline Mode & Exp\cite{Amsler:2008zzb,phiphip,phiphibr}&  \multicolumn{5}{c}{This work} \\
\cline{3-7}
                        &               &   NLO        &   NLO$^{eff} $   &   NLO$^{eff}(45^\circ)$     & NLO$^{eff}(60^\circ)$ & NLO$^{eff}(75^\circ)$    \\
\hline
$B_s \to \rho^0 \bar K^{*0}$ & $<767$& 0.6&0.8&0.7&$0.7$&0.7\\
$B_s\to \rho^+K^{*-}$ &  & 23.6&21.0&20.7&$20.6$&20.6\\

$B_s \to K^{*-} K^{*+}$& & 13.4&12.8&11.0&$10.4$&9.8\\
$B_s \to K^{*0} \bar{K}^{*0}$& $<1681$&15.0&13.1&10.6&$9.8$&9.1\\

$B_s \to \phi\phi$& $24.0 \pm 8.9$ & 22.1&18.7&12.1&$10.0$&7.9\\
\hline
$A_{CP}(\rho^0\bar K^{*0})$     &           & 66.8&56.4&60.8&$56.8$ &50.0 \\
$A_{CP}(\rho^+ K^{*-})$  &         & -10.1&-10.8&-11.3&$-9.7$&-7.2\\
$A_{CP}(K^{*-} K^{*+})$            &     & 20.1&17.9&26.3&$26.4$& 24.6\\
$A_{CP}(K^{*0} \bar K^{*0})$         &       &0&0&0&$0$&0  \\
$A_{CP}(\phi\phi)$ &      &0&0&0&$0$&0  \\
\hline
$f_L(\rho^0 \bar K^{*0})$ &$$& 80&84&79&$77$&76\\
$f_L(\rho^+K^{*-})$ &$$& 96&96&96&$95$&95\\

$f_L(K^{*-} K^{*+})$ &$$ &72&71&54&$48$&43\\
$f_L(K^{*0} \bar{K}^{*0})$ &$$ &76&72&50&$41$&32\\

$f_L(\phi\phi)$ &$34.8\pm 4.1\pm 2.1$ &71&65&50&$42$&31\\
\hline\hline
\end{tabular}}
\end{center}
\end{table}
%%%%%%%%%%%%%%%%%%%%%%%%
\begin{table}[tb]
\caption{The comparisons in theoretical methods in $B_s\to VV$ decays. The central values are obtained with $\mu_g^a$=0.52GeV and $\theta^{a} = 60^{\circ}$. The first error in our predictions arises from the varying for $\mu_{scale}=1.4\sim 1.6$ GeV, the second one stems from the shape parameters of light mesons.}\label{tab:br44}
 \begin{center}\scriptsize{
\begin{tabular}{c|c|c|c|ccc}
\hline \hline Mode & Exp~\cite{Amsler:2008zzb,phiphip,phiphibr}&  QCDF~\cite{Cheng:2009mu} & pQCD~\cite{Ali:2007ff} &\multicolumn{3}{c}{This work} \\
\cline{5-7}
                        &                  &           &         & LO        & NLO &NLO($a^{eff},\theta^{a}$)    \\
\hline
$B_s \to \rho^0 \bar K^{*0}$ & $<767$& $1.3^{+2.0+1.7}_{-0.6-0.3}$& $0.33^{+0.09+0.14+0.00}_{-0.07-0.09-0.01}$ &0.2&0.6&$1.0^{+0.3+0.3}_{-0.2-0.2}$\\
$B_s\to \rho^+K^{*-}$ &  &  $21.6^{+1.3+0.9}_{-2.8-1.5}$& $20.9^{+8.2+1.4+1.2}_{-6.2-1.4-1.1}$ &22.3&23.6&$21.0^{+13.4+2.6}_{-6.2-1.8}$\\

$B_s \to K^{*-} K^{*+}$& & $7.6^{+1.0+2.3}_{-1.0-1.8}$& $6.7^{+1.5+3.4+0.5}_{-1.2-1.4-0.2}$&10.3&13.4&$10.4^{+3.0+2.7}_{-2.5-1.6}$\\
$B_s \to K^{*0} \bar{K}^{*0}$& $<1681$&$6.6^{+1.1+1.9}_{-1.4-1.7}$& $7.8^{+1.9+3.8+0.0}_{-1.5-2.2-0.0}$&10.9&15.0&$9.8^{+3.1+2.5}_{-2.2-2.1}$\\

$B_s \to \phi\phi$& $24.0 \pm 8.9$ & $16.7^{+2.6+11.3}_{-2.1-8.8}$& $35.3^{+8.3+16.7+0.0}_{-6.9-10.2-0.0}$&18.9&22.1&$10.0^{+2.9+3.1}_{-2.0-2.3}$\\

$B_s \to \rho^+\rho^-$& & $0.68^{+0.04+0.73}_{-0.04-0.53}$& $1.0^{+0.2+0.3+0.0}_{-0.2-0.2-0.0}$&0.56&0.70&$0.70^{+0.01+0.05}_{-0.01-0.05}$\\
$B_s \to \rho^0\rho^0$& $<320$& $0.34^{+0.02+0.36}_{-0.02-0.26}$& $0.51^{+0.12+0.17+0.01}_{-0.11-0.10-0.01}$&0.28&0.35&$0.35^{+0.01+0.02}_{-0.01-0.02}$\\
\hline
$A_{CP}(\rho^0\bar K^{*0})$     &           & $46^{+15+10}_{-17-25}$& $61.8^{+3.2+17.1+4.4}_{-4.7-22.8-2.3}$ &52.2&66.8&$56.8^{+1.0+3.0}_{-0.5-2.9}$  \\
$A_{CP}(\rho^+ K^{*-})$  &         & $-11^{+1+4}_{-1-1}$& $-8.2^{+1.0+1.2+0.4}_{-1.2-1.7-1.1}$ &-10.0&-10.1&$-9.7^{+3.5+1.3}_{-3.0-1.3}$\\
$A_{CP}(K^{*-} K^{*+})$            &      & $21^{+1+2}_{-2-4}$&  $9.3^{+0.4+3.3+0.3}_{-0.7-3.6-0.2}$ &16.1&20.1&$26.4^{+2.4+2.1}_{-2.5-2.5}$ \\
$A_{CP}(K^{*0} \bar K^{*0})$         &       & $0.4^{+0.8+0.6}_{-0.5-0.4}$& $0$ &0&0&$0^{+0+0}_{-0-0}$  \\

$A_{CP}(\phi\phi)$ & & $0.2^{+0.4+0.5}_{-0.3-0.2}$ & $0$&$0^{+0+0}_{-0-0}$\\
$A_{CP}({\rho}^{+}\rho^-)$     &      & $0$& $-2.1^{+0.2+1.7+0.1}_{-0.1-1.3-0.1}$ &5.8&5.0&$5.0^{+1.2+0.4}_{-2.5-0.4}$\\
$A_{CP}({\rho}^{0}\rho^0)$     &      & $0$& $-2.1^{+0.2+1.7+0.1}_{-0.1-1.3-0.1}$ &5.8&5.0&$5.0^{+1.2+0.4}_{-2.5-0.4}$\\
\hline
$f_L(\rho^0 \bar K^{*0})$&$$& $90^{+4+3}_{-5-23}$&$45.5^{+0.4+6.9+0.6}_{-0.3-4.3-0.9}$ &73&80&$77^{+2+0}_{-1-0}$\\
$f_L(\rho^+K^{*-})$ &$$& $92^{+1+1}_{-2-3}$& $93.7^{+0.1+0.2+0.0}_{-0.2-0.3-0.2}$ &96&96&$95^{+1+0}_{-0-0}$\\

$f_L(K^{*-} K^{*+})$ &$$ &$52^{+3+20}_{-5-21}$& $43.8^{+5.1+2.1+3.7}_{-4.0-2.3-1.5}$ &66&72&$48^{+4+2}_{-4-2}$\\
$f_L(K^{*0} \bar{K}^{*0})$ &$$ &$56^{+4+22}_{-7-26}$ &$49.7^{+5.7+0.6+0.0}_{-4.8-3.8-0.0}$&69&76&$41^{+3+1}_{-3-1}$\\

$f_L(\phi\phi)$ &$34.8\pm 4.1\pm 2.1$ &$36^{+3+23}_{-4-18}$&$61.9^{+3.6+2.5+0.0}_{-3.2-3.3-0.0}$ &65&71&$42^{+3+2}_{-3-2}$\\

$f_L(\rho^+\rho^-)$ &$$ & $100$&$\sim$ 100 &$\sim$ 100 &$\sim$ 100 & $\sim$ 100 \\
$f_L(\rho^0\rho^0)$ &$$ & $100$&$\sim$ 100 &$\sim$ 100 &$\sim$ 100 & $\sim$ 100 \\
\hline\hline
\end{tabular}}
\end{center}
\end{table}
%%%%%%%%%%%%%%%%%%%%%%%%%%%

\subsubsection*{1. $B_s \to \rho^0 \bar K^{*0}, \rho^+ \bar K^{*-}$}

These two decay channels are color-suppressed and tree-dominated, respectively. From Tables~\ref{tab:br44-1} and \ref{tab:br44}, it is noted that the decay mode $B_s \to \rho^+ \bar K^{*-}$ has sizable branching ratio of order $21\times 10^{-6}$, moderate direct CP asymmetry, and very big longitudinal polarization, which are all in good agreement with the predictions in QCDF~\cite{Cheng:2009mu} and pQCD~\cite{Ali:2007ff} methods. As for the color-suppressed mode $B_s \to \rho^0 \bar K^{*0}$, it has remarkable direct CP asymmetry of $56.8\%$ and longitudinal polarization of $76\%$, which are also well consistent with the other predictions. Moreover, the effects of different strong phases on branching ratios, direct CP asymmetries and longitudinal polarizations are insignificant in these two decays.

\subsubsection*{2. $B_s \to K^{*+} K^{*-}, K^{*0} \bar K^{*0},  \phi\phi$}

As for these three decay modes, they are all penguin-dominated and their branching ratios are at the order of $10\times 10^{-6}$. As for the direct CP asymmetry, the ones of $B_s\to K^{*0} \bar K^{*0}$ and $B_s\to \phi\phi$ are zero, while there is a big direct CP asymmetry in $B_s\to K^{*+} K^{*-}$ channel. Moreover, recent data from the CDF Collaboration favors a huge transverse polarizations, which is denoted by $f_T\equiv1-f_L$ in $B_s\to \phi\phi$ decay; our prediction is consistent with the data and also agrees with the one in QCDF method~\cite{Cheng:2009mu}. In addition, the predictions of the longitudinal polarization fraction, which are less than $50\%$ in $B_s \to K^{*+} K^{*-}$ and $B_s \to K^{*0} \bar K^{*0} $ are similar to the one in $\phi\phi$ mode.

One important point should be addressed is that the annihilation contributions with a strong phase have remarkable effects on the branching ratios, direct CP asymmetries and longitudinal polarizations in these decays. Especially to be mentioned,the effects of effective Wilson coefficients and strong phase decrease the branching ratio in $B_s\to \phi\phi$ mode a lot. As a result, the prediction of $Br(B_s\to \phi\phi)$ is much smaller than the current data and also that in the other theoretical methods. This discrepancy is due to the fact that we have taken a bigger gluon infrared cut-off in annihilation diagram, $\widetilde{\mu}_g =520$GeV, following what we have done in our previous work for the $B$ decays~\cite{Su:2010vt}. This results in a smaller branching ratio in $B_s\to \phi\phi$ decay since the bigger $\widetilde{\mu}_g$ we take, the smaller contributions to the decay amplitude.

It is known that we can relate these three decay modes to $B_d\to K^{*-} \rho^+$, $K^{*0} \rho^0$ and $\phi K^{*0}$ by SU(3) symmetry. Although our prediction of $Br(B_s\to \phi\phi)=10.0^{+2.9}_{-2.1}$, is much smaller than the data, it satisfies the SU(3) symmetry through the following ratio:
\begin{eqnarray}
R_{Br} =  \frac{Br(B_s\to \phi\phi)}{Br(B_d\to \phi K^{*0})} \approx 1.08.
\end{eqnarray}
Moreover, the predictions of huge transverse polarizations are reasonable in these $B_s$ decays since the three penguin-dominated B decays also have  huge transverse polarizations. As the data still has a big uncertainty, it is expected that more precise experimental measurements will be helpful to clarify such an issue.

\subsubsection*{3. $B_s\to \rho^+\rho^-, \rho^0\rho^0$}

Now we proceed to discuss the final two decay modes, which involve only the annihilation contributions, thus it is meaningless to discuss the phase influence which we shall skip in these decays. Since there is a $\sqrt{2}$ factor between the amplitudes of these two decays, the predictions for the direct CP asymmetry, longitudinal polarization are the same but with a factor of 2 difference in branching ratios. Our predictions for the branching ratios are smaller than those in pQCD~\cite{Ali:2007ff} but consistent with QCDF~\cite{Cheng:2009mu} method.

\section{Conclusions}\label{sec:conc}

Based on the approximate six-quark operator effective Hamiltonian derived from QCD, the naive factorization approach has been naturally applied to evaluate the hadronic matrix elements for charmless two body B-meson decays. It has been shown that, when considering annihilation contributions and extra strong phase effects, our framework provides a simple way to evaluate the hadronic matrix elements of two-body $B_s$ decays.

For $B_s\to PP, PV$ final states, our predictions for the branching ratios and CP asymmetries are roughly consistent with those of the other theoretical methods within their respective uncertainties, once the effective Wilson coefficients and annihilation amplitude with small strong phase~($\theta^a=5^\circ$) are adopted. The exception here is the branching ratio of $B_s\to \pi^+K^-$ mode, which is a little bigger than the data, but it is interesting to note that the prediction is well consistent with the ones in pQCD method~\cite{Ali:2007ff}. As the current data on the branching ratio has large uncertainties in this mode, more precise experimental data are expected to further test our prediction. In the $B_s\to VV$ decay modes, it is noticed that there are huge transverse polarization fractions in penguin-dominated decays in our framework when considering annihilation contributions with a large strong phase~($\theta^a=60^\circ$). Moreover, the prediction for the branching ratio in $B_s\to \phi\phi$ is below the current data but can be explained by the SU(3) symmetry in comparison with the $B$ decays.

Another important point should be addressed is that the method developed in this paper allows us to calculate the relevant transition form factors. Our predictions for $B_s$ to light mesons form factors are consistent with the results of light-cone QCD sum rules and QCD sum rules. In this sense, we can say that our framework is reasonable from both the theoretical considerations and the phenomenological applications to the hadronic bottom meson decays.

Generally, it is observed that the predictions for the branching ratios of the tree-dominated $B_s$ decays are in good agreements among different theoretical methods, while there are big discrepancies in the color-suppressed, penguin-dominated and annihilation $B_s$ decays. QCDF method~\cite{Cheng:2009mu} favors big color-suppressed and penguin-dominated contributions, while pQCD method~\cite{BsPV} prefers big annihilation contributions. The predictions from our method stand between those results given by the other two methods. It is expected that the future more precise experimental datas from LHC-b and Super B factories will provide a better test and clarify the relevant important issues.

\section*{\textbf{Acknowledgements: }}

This work was supported in part by the National Science Foundation
of China (NSFC) under the grant \# 10821504, 10975170,11047165 and the Project of
Knowledge Innovation Program (PKIP) of Chinese Academy of Sciences.

\section*{Appendix: Decay amplitudes of $B_s\to PP, PV, VV$ modes}\label{sec:Amplitude}
\def\theequation{A.\arabic{equation}}
\def\thesubsection{A}
\setcounter{equation}{0}

As discussed in Section~\ref{sec:sqeh}, the QCD factorization approach with six-quark operator effective Hamiltonian enables us to evaluate all the hadronic matrix elements in two-body hadronic B-meson decays. Here we list only the results for the various decay amplitudes expressed in terms of different topological amplitudes, which could be found in the appendix of our previous paper~\cite{Su:2010vt}.

The detailed calculations of the hadronic matrix elements for $B_s\to PP$ decays could be found in our previous paper~\cite{Su:2008mc,Su:2010vt}. As for the decay amplitudes and the hadronic matrix elements for $B_s\to PV, VV$ decays, we shall list them one by one as follows. Firstly, for $B_s\to \pi\rho$ decay channels, we have
\begin{eqnarray}
  A(B_s^0\to \rho^+\pi^-)&=&-V_{us}V^*_{ub}E^{\pi\rho}(B_s)+V_{ts}V^*_{tb}[2P_{A}^{\pi\rho}(B_s)
  +\frac{1}{3}P_{EW}^{A\pi\rho}(B_s)],\nonumber\\[0.2cm]
  A(B_s^0\to \pi^+\rho^-)&=&-V_{us}V^*_{ub}E^{\pi\pi}(B_s)+V_{ts}V^*_{tb}[2P_{A}^{\rho\pi}(B_s)
  +\frac{1}{3}P_{EW}^{A\rho\pi}(B_s)],\nonumber\\[0.2cm]
  A(B_s^0\to \pi^0\rho^0)&=&\frac{1}{2}(A(B_s^0\to \rho^+\pi^-)+A(B_s^0\to \pi^+\rho^-)).
\end{eqnarray}

For $B_s\to \pi K^*$ decay channels, the amplitudes are
\begin{eqnarray}
  A(B_s^0\to \pi^+ K^{*-})&=&V_{td}V^*_{tb}[P^{\bar{K}^*\pi}(B_s)+\frac{2}{3}P_{EW}^{C\bar{K}^*\pi}(B_s)
  +P_{E}^{\bar{K}^*\pi}(B_s)
  -\frac{1}{3}P_{EW}^{E\bar{K}^*\pi}(B_s)]\nonumber\\&&-V_{us}V^*_{ub}T^{\bar{K}^*\pi}(B_s),\nonumber\\[0.2cm]
  A(B_s^0\to \pi^0 \bar K^{*0})&=&-\frac{1}{\sqrt{2}}\{V_{td}V^*_{tb}[P^{\bar{K}^*\pi}(B_s)-P_{EW}^{\bar{K}^*\pi}(B_s)
  -\frac{1}{3}P_{EW}^{C\bar{K}^*\pi}(B_s)+P_{E}^{\bar{K}^*\pi}(B_s)\nonumber\\
  &&-\frac{1}{3}P_{EW}^{E\bar{K}^*\pi}(B_s)]
  +V_{us}V^*_{ub}C^{\bar{K}^*\pi}(B_s)\}.
\end{eqnarray}

For $B_s\to \rho K $ decay channels, we have
\begin{eqnarray}
  A(B_s^0\to \rho^+ K^-)&=&V_{td}V^*_{tb}[P^{\bar{K}\rho}(B_s)+\frac{2}{3}P_{EW}^{C\bar{K}\rho}(B_s)+P_{E}^{\bar{K}\rho}(B_s)
  -\frac{1}{3}P_{EW}^{E\bar{K}\rho}(B_s)]\nonumber\\&&-V_{us}V^*_{ub}T^{\bar{K}\rho}(B_s),\nonumber\\[0.2cm]
  A(B_s^0\to \rho^0 \bar K^0)&=&-\frac{1}{\sqrt{2}}\{V_{td}V^*_{tb}[P^{\bar{K}\rho}(B_s)-P_{EW}^{\bar{K}\rho}(B_s)
  -\frac{1}{3}P_{EW}^{C\bar{K}\rho}(B_s)+P_{E}^{\bar{K}\rho}(B_s)\nonumber\\
  &&-\frac{1}{3}P_{EW}^{E\bar{K}\rho}(B_s)]
  +V_{us}V^*_{ub}C^{\bar{K}\rho}(B_s)\}.
\end{eqnarray}

For $B_s\to K^* K $ decay channels, it is given by
\begin{eqnarray}
  A(B_s^0\to K^+K^{*-})&=&-V_{ts}V^*_{tb}[P^{{K}^*K}(B_s)+\frac{2}{3}P_{EW}^{C{K}^*K}(B_s)+P_E^{{K}^*K}(B_s)+
  P_{A}^{{K}^*K}(B_s)\nonumber\\&&+P_{A}^{K{K}^*}(B_s)+\frac{2}{3}P_{EW}^{AK{K}^*}(B_s)
  -\frac{1}{3}P_{EW}^{A{K}^*K}(B_s)
  -\frac{1}{3}P_{EW}^{E{K}^*K}(B_s)]\nonumber\\&&+V_{us}V^*_{ub}[T^{{K}^*K}(B_s)
  +E^{{K}^*K}(B_s)],
\end{eqnarray}
\begin{eqnarray}
  A(B_s^0\to K^{0}\bar{K}^{*0})&=&-V_{ts}V^*_{tb}[P^{\bar{K}^*K}(B_s)-\frac{1}{3}P_{EW}^{C\bar{K}^*K}(B_s)
  +P_E^{\bar{K}^*K}(B_s)+
  P_{A}^{\bar{K}^*K}(B_s)\nonumber\\&&+P_{A}^{K\bar{K}^*}(B_s)-\frac{1}{3}P_{EW}^{AK\bar{K}^*}(B_s)
  -\frac{1}{3}P_{EW}^{A\bar{K}^*K}(B_s)
  -\frac{1}{3}P_{EW}^{E\bar{K}^*K}(B_s)],\nonumber\\[0.2cm]
  A(B_s^0\to K^{*+}K^-)&=&-V_{ts}V^*_{tb}[P^{{K}K^*}(B_s)+\frac{2}{3}P_{EW}^{C{K}K^*}(B_s)+P_E^{{K}K^*}(B_s)+
  P_{A}^{{K}K^*}(B_s)\nonumber\\&&+P_{A}^{K^*{K}}(B_s)+\frac{2}{3}P_{EW}^{AK^*{K}}(B_s)
  -\frac{1}{3}P_{EW}^{A{K}K^*}(B_s)
  -\frac{1}{3}P_{EW}^{E{K}K^*}(B_s)]\nonumber\\&&+V_{us}V^*_{ub}[T^{{K}K^*}(B_s)
  +E^{{K}K^*}(B_s)],\nonumber\\[0.2cm]
  A(B_s^0\to K^{*0}\bar{K}^{0})&=&-V_{ts}V^*_{tb}[P^{\bar{K}K^*}(B_s)-\frac{1}{3}P_{EW}^{C\bar{K}K^*}(B_s)
  +P_E^{\bar{K}K^*}(B_s)+
  P_{A}^{\bar{K}K^*}(B_s)\nonumber\\&&+P_{A}^{K^*\bar{K}}(B_s)-\frac{1}{3}P_{EW}^{AK^*\bar{K}}(B_s)
  -\frac{1}{3}P_{EW}^{A\bar{K}K^*}(B_s)
  -\frac{1}{3}P_{EW}^{E\bar{K}K^*}(B_s)].
\end{eqnarray}

As for the $B_s\to VV$ decays, since there are three kinds of polarizations for a vector meson, namely, longitudinal~($L$), perpendicular~($\bot$) and parallel~($\parallel$), the amplitudes are also characterized by the polarization states of these two vector mesons. Here, we only list the longitudinal ones and the other ones are similar. They are given by
\begin{eqnarray}
  A(B_s^0\to \rho^+\rho^-)&=&-V_{us}V^*_{ub}E^{\rho\rho}(B_s)+V_{ts}V^*_{tb}[2P_{A}^{\rho\rho}(B_s)
  +\frac{1}{3}P_{EW}^{A\rho\rho}(B_s)],\nonumber\\[0.2cm]
  A(B_s^0\to \rho^0\rho^0)&=&\frac{1}{\sqrt{2}}A(B_s^0\to \rho^+\rho^-).
\end{eqnarray}
for $B_s\to \rho \rho$ decay channels, and
\begin{eqnarray}
  A(B_s^0\to \rho^+ K^{*-})&=&V_{td}V^*_{tb}[P^{\bar{K}^*\rho}(B_s)+\frac{2}{3}P_{EW}^{C\bar{K}^*\rho}(B_s)
  +P_{E}^{\bar{K}^*\rho}(B_s)
  -\frac{1}{3}P_{EW}^{E\bar{K}^*\rho}(B_s)]\nonumber\\
  &&-V_{us}V^*_{ub}T^{\bar{K}^*\rho}(B_s),\nonumber\\[0.2cm]
  A(B_s^0\to \rho^0 \bar K^{*0})&=&-\frac{1}{\sqrt{2}}\{V_{td}V^*_{tb}[P^{\bar{K}^*\rho}(B_s)-P_{EW}^{\bar{K}^*\rho}(B_s)
  -\frac{1}{3}P_{EW}^{C\bar{K}^*\rho}(B_s)
  +P_{E}^{\bar{K}^*\rho}(B_s)\nonumber\\&&-\frac{1}{3}P_{EW}^{E\bar{K}^*\rho}(B_s)]
  +V_{us}V^*_{ub}C^{\bar{K}^*\rho}(B_s)\}.
\end{eqnarray}
for $B_s\to \rho K^* $ decay channels, and
\begin{eqnarray}
  A(B_s^0\to K^{*+}K^{*-})&=&-V_{ts}V^*_{tb}[P^{{K}^*K^*}(B_s)+\frac{2}{3}P_{EW}^{C{K}^*K^*}(B_s)+P_E^{{K}^*K^*}(B_s)+
  P_{A}^{{K}^*K^*}(B_s)\nonumber\\&&+P_{A}^{K^*{K}^*}(B_s)+\frac{2}{3}P_{EW}^{AK^*{K}^*}(B_s)
  -\frac{1}{3}P_{EW}^{A{K}^*K^*}(B_s)
  -\frac{1}{3}P_{EW}^{E{K}^*K^*}(B_s)]\nonumber\\&&+V_{us}V^*_{ub}[T^{{K}^*K^*}(B_s)
  +E^{{K}^*K^*}(B_s)],\nonumber\\[0.2cm]
  A(B_s^0\to K^{*0}K^{*0})&=&-V_{ts}V^*_{tb}[P^{\bar{K}^*K^*}(B_s)-\frac{1}{3}P_{EW}^{C\bar{K}^*K^*}(B_s)
  +P_E^{\bar{K}^*K^*}(B_s)+
  P_{A}^{\bar{K}^*K^*}(B_s)\nonumber\\&&+P_{A}^{K^*\bar{K}^*}(B_s)-\frac{1}{3}P_{EW}^{AK^*\bar{K}^*}(B_s)
  -\frac{1}{3} P_{EW}^{A\bar{K}^*K^*}(B_s)
  -\frac{1}{3}P_{EW}^{E\bar{K}^*K^*}(B_s)].
\end{eqnarray}
for $B_s\to K^* K^* $ decay channels, and
\begin{eqnarray}
  A(B_s^0\to\phi\phi)&=&-V_{ts}V^*_{tb}[P^{\phi\phi}(B_s)+P^{C\phi\phi}(B_s)
  -\frac{1}{3}P_{EW}^{\phi\phi}(B_s)-\frac{1}{3}P_{EW}^{C\phi\phi}(B_s)
  +P_E^{\phi\phi}(B_s)\nonumber\\
  &&+P_A^{\phi\phi}(B_s)-\frac{1}{3}P_{EW}^{A\phi\phi}(B_s)-\frac{1}{3}P_{EW}^{E\phi\phi}(B_s)].
\end{eqnarray}
for $B_s\to \phi  \phi $ decay channel.
%%%%%%%%%%%%%%%

\end{document}